\pdfoutput=1
\documentclass[aps,prd,amsmath,floats,floatfix, twocolumn,
superscriptaddress,nofootinbib,showpacs]{revtex4-1}

\usepackage[T1]{fontenc}
\usepackage[utf8]{inputenc}
\usepackage{lmodern}
\usepackage{tensor}
\usepackage{verbatim}
\usepackage{siunitx}
\usepackage[caption=false]{subfig}
\usepackage{wasysym}
\usepackage[english]{babel}
\usepackage{booktabs}
\usepackage{mathrsfs}
\usepackage{multirow}

\usepackage[dvipsnames, usenames]{xcolor}
\definecolor{linkcolor}{rgb}{0.0,0.3,0.5}
\usepackage[hypertexnames=false, unicode, colorlinks=true, linkcolor=linkcolor,
citecolor=linkcolor, filecolor=linkcolor,urlcolor=linkcolor,
pdfusetitle]{hyperref}

\usepackage[all]{hypcap}
\usepackage{graphicx}
\usepackage{xspace}
\usepackage{amssymb}
\usepackage[normalem]{ulem} %for \sout
\usepackage{bm} % boldmath
\usepackage{enumitem,amssymb}
\usepackage{mathtools} 

\usepackage{microtype}

\usepackage[english]{babel}
\usepackage{blindtext}
\usepackage{tikz}
\usetikzlibrary{decorations.pathmorphing}
\usetikzlibrary{decorations.markings}
\usepackage{orcidlink}

\graphicspath{%
  {figs/}%
}

\DeclareMathAlphabet{\mathpzc}{OT1}{pzc}{m}{it}

\newlist{todolist}{itemize}{2}
\setlist[todolist]{label=$\square$}

\allowdisplaybreaks

\begin{document}

\title{Relativistic excitation of  compact stars}

\newcommand{\Cornell}{\affiliation{Cornell Center for Astrophysics
    and Planetary Science, Cornell University, Ithaca, New York 14853, USA}}
\newcommand\CornellPhys{\affiliation{Department of Physics, Cornell
    University, Ithaca, New York 14853, USA}}
\newcommand\Caltech{\affiliation{TAPIR 350-17, California Institute of
    Technology, 1200 E California Boulevard, Pasadena, CA 91125, USA}}
\newcommand{\AEI}{\affiliation{Max Planck Institute for Gravitational Physics
    (Albert Einstein Institute), Am M\"uhlenberg 1, Potsdam 14476, Germany}} %
\newcommand{\UMassD}{\affiliation{Department of Mathematics,
    Center for Scientific Computing and Visualization Research,
    University of Massachusetts, Dartmouth, MA 02747, USA}}
\newcommand\Olemiss{\affiliation{Department of Physics and Astronomy,
    The University of Mississippi, University, MS 38677, USA}}
\newcommand{\Bham}{\affiliation{School of Physics and Astronomy and Institute
    for Gravitational Wave Astronomy, University of Birmingham, Birmingham, B15
    2TT, UK}}
\newcommand{\Perimeter}{\affiliation{Perimeter Institute for Theoretical Physics, Waterloo, ON N2L2Y5, Canada}}
\newcommand{\UGuelph}{\affiliation{University of Guelph, Guelph, Ontario N1G 2W1, Canada}}
\newcommand{\Tsinghua}{\affiliation{Department of Astronomy, Tsinghua University, Beijing 100084, China}}

\author{Zhiqiang Miao\orcidlink{0000-0003-1197-3329}}
\affiliation{Tsung-Dao Lee Institute, Shanghai Jiao Tong University, Shanghai, 1 Lisuo Road, 201210, China}
\author{Xuefeng Feng\orcidlink{0000-0002-3195-6924}}
\affiliation{Fudan Center for Mathematics and Interdisciplinary Study, Fudan University, Shanghai, 200433, China}
\affiliation{Shanghai Institute for Mathematics and Interdisciplinary Sciences (SIMIS), Shanghai, 200433, China}

\author{Zhen Pan\orcidlink{0000-0001-9608-009X}}
\affiliation{Tsung-Dao Lee Institute, Shanghai Jiao Tong University, Shanghai, 1 Lisuo Road, 201210, China}
\affiliation{School of Physics \& Astronomy, Shanghai Jiao-Tong University, Shanghai, 800 Dongchuan Road, 200240, China}

\author{Huan Yang\orcidlink{0000-0002-9965-3030}}
\email{hyangdoa@tsinghua.edu.cn}
\Tsinghua

\date{\today}

%==========================================================================
\begin{abstract}
In this work, we study the excitation of a compact star under the influence of external gravitational driving in the relativistic regime. Using a model setup in which a wave with constant frequency is injected from past null infinity and scattered by the star to future null infinity, we show that the scattering coefficient encodes rich information of the star. For example, the analytical structure of the scattering coefficient implies that the decay rate of a mode generally plays the role of ``star excitation factor'', similar to the ``black hole excitation factor'' previously defined for describing black hole mode excitations. With this star excitation factor we derive the relativistic transient mode excitation as a binary system crosses a generic mode resonance of a companion star during the inspiral stage. This application is useful because previous description of resonant mode excitation of stars still relies on the mode and driving force decomposition based on the Newtonian formalism. The relativistic transient mode energy may differ from the Newtonian prediction by one order of magnitude. In addition, we show that the scattering phase is intimately related to the total energy of spacetime and matter under the driving of a steady input wave from infinity. We also derive the relevant tidal energy of a star under steady driving and compare that with the dynamic tide formula. We estimate that the difference may lead to $\mathcal{O}(0.5)$ radian phase modulation in the late stage of the binary neutron star inspiral waveform.
\end{abstract}

\maketitle

% \textcolor{PineGreen}{
% \begin{todolist}
%    % \item[\checkmark] 
%     % \item There is an overall minus sign in my code!!!! I need to fix the expressions in the paper eventually.
%     % \item retrograde mode is not an accurate term \red{Really? I'm not sure now}
%     % \item gauge effects in~\cite{Spiers:2023cip}
%     % \item  \red{Asymptotic behavior}
% \end{todolist}
% }
%\done
%==========================================================================

\section{Introduction}
\label{sec:introduction}

Compact neutron star binaries are important sources for ground-based gravitational wave detection. Compared to black hole binaries, neutron stars contribute additional matter effects to the gravitational waveform, including (dynamic) tides~\cite{Lai:1993pa,Flanagan:2007ix,Steinhoff:2016rfi}, mode resonances~\cite{Lai:1993di,Lai:2006pr,Yu:2017cxe,Yang:2018bzx,Yang:2019kmf,Pan:2020tht,Poisson:2020eki,Ma:2020oni,Lau:2020bfq,Kwon:2024zyg,Kwon:2025zbc,Miao:2023jqe,Ho:1998hq,Xu:2017hqo}, and various post-merger signatures~\cite{Shibata:2006nm,Kiuchi:2009jt,Bauswein:2011tp,Bauswein:2015vxa,Palenzuela:2015dqa,Yang:2017xlf,Baiotti:2016qnr,Paschalidis:2016vmz}. The post-merger stage usually takes place in the kilohertz range, so the associated gravitational wave emission is normally only observable with third-generation gravitational wave detectors~\cite{Miao:2017qot,Martynov:2019gvu,Zhang:2022yab}. In contrast, both dynamic tides and mode resonances occur during the inspiral stage with lower frequency. Their detection may inform us about the equation-of-state and the internal structure of neutron stars~\cite{Hinderer:2007mb,Flanagan:2007ix,Hinderer:2016eia,LIGOScientific:2017vwq}.

Both effects are excited because of the gravitational force from the companion star. The current treatment of neutron star excitations assumes an effective model with one or more harmonic oscillators, and the external gravitational field is directly coupled to the oscillators~\cite{Flanagan:2007ix,Steinhoff:2016rfi,Lai:1993di}. Such a physical picture is justified in the Newtonian regime, as the fluid equation of motion within a star can be formally decomposed into a mode basis~\cite{Lai:1993di}: %e.g. as discussed in~\cite{}. 
%However, for a compact, relativistic star, it is not necessarily clear whether the mode projection and decomposition procedure is still valid, as 
\begin{align}\label{eq:newton}
\ddot{a}_n +\omega^2_n {a}_n =\langle f, \psi_n \rangle
\end{align}
with $\omega_n$ being the eigenfrequency of the mode, $\psi_n$ being the mode wavefunction, $f$ being the external gravitational force, ${a}_n$ being the mode amplitude, $\langle \rangle$ being an inner product such that different modes are orthogonal to each other, and $\langle f, \psi_n \rangle$ being the overlap integral {normalized by $\langle \psi_n, \psi_n \rangle$}.

However, for a compact star in the fully relativistic regime, the local mass element of the star is still driven by the local gravitational perturbation, whereas the gravitational perturbation is both influenced by the external gravitational field and the fluid. In other words, it is no longer a good approximation to assume local gravitational perturbations by taking the values as if the star were not present. 
The gravitational perturbations and fluid
perturbations must evolve and be computed consistently. Because the gravitational sector and the fluid sector are coupled, it is not clear whether the mode excitation can still be computed using the projection of external force. In fact, since generic modes have non-zero decay rates, they no longer form a complete set~\cite{Kokkotas:1999bd} and their wavefunctions are not orthogonal to each other using the vanilla inner product in Eq.~(\ref{eq:newton}).

Therefore, it is necessary to construct a formalism that applies in the fully relativistic regime. Earlier studies~\cite{Pitre:2023xsr} extended the equilibrium tide calculation to a time-dependent regime by computing new tidal coefficients, but a complete theory describing the tidal dynamics of compact stars is still needed. In this work, we consider the excitation of a star with constant-frequency gravitational waves injected from past null infinity and scattered back to future null infinity. This setup describes generic time-dependent tidal drivings in the linear (amplitude) regime, as any time-dependent tidal driving can be Fourier decomposed into various frequency components. 
In addition, it can be shown that (see Sec.~\ref{sec:scat}) this wave-scattering problem is dual to the problem with a point mass orbiting around the star, which can be considered as the extremal-mass-ratio limit of a binary problem \footnote{{In a series of works in \cite{Gualtieri:2001cm,Pons:2001xs,Berti:2002ry}, the star $+$ point mass scenario was applied to numerically compute the contribution of stellar excitations to the gravitational-wave flux and the waveform.}}.
It is different but related to the treatment in \cite{Pitre:2023xsr} using dynamical Love numbers, which may be defined using the frequency expansion of our results.

With this scattering formalism we compute the scattering coefficient defined using the (complex) amplitudes of outgoing and ingoing waves at infinity, which encodes information about the tidal response of the star~\footnote{We note that the scattering formalism is also applied in the world-line effective field theory to study tide effects, in which the scattering coefficients can be related to the love number and dissipation number~\cite{Creci:2021rkz,Saketh:2024juq,Ivanov:2022qqt}.}.
The poles of the scattering coefficient correspond to quasinormal modes of the compact star. We argue that the poles and zeros of the scattering coefficient form conjugate pairs $\omega_{\rm R} \pm i \omega_{\rm I}$ in the complex frequency plane~\footnote{This argument strictly applies only to conservative systems and does not hold exactly for dissipative systems. Neutron stars do have dissipation due to viscosity, but it is typically very small~\cite{Lai:1993di,Ripley:2023qxo}; for simplicity, we neglect neutron star dissipation in this work.}, which implies that the residue of a pole is well approximated by the imaginary part $\omega_{\rm I}$ of the pole, if $\omega_{\rm I} \ll \omega_{\rm R}$. This observation provides remarkable simplification in understanding the black hole excitation factor, as detailed in Sec.~\ref{sec:scat}.

For a time-dependent driving field, the time-domain response of the star can be obtained by Fourier transforming its frequency-domain signal. We show that this naturally leads to transient excitations of quasinormal modes of the star, with information purely from the scattering coefficient, so that a mode projection procedure is no longer required.
We further derive the transient energy of the modes in this relativistic regime, which agrees with its counterpart~\cite{Lai:1993di} in the Newtonian limit. The transient mode energy is generally proportional to the residue or the decay rate $\omega_{\rm I}$ of the mode, which plays the role of an overlap integral between the mode wave function and the external gravitational field in Newtonian theory [Eq.~(\ref{eq:newton})]. The residue is also analogous to the black hole excitation factor that describes the degree of black hole mode excitations in black hole ringdowns~\cite{Leaver:1986gd}.

For a star under a steady external driving field, we show that the energy of gravitational and matter perturbations can be entirely determined by the amplitude of ingoing gravitational waves and the scattering angle
of the outgoing wave. This is also numerically shown using the Hamiltonian of star perturbations derived in Refs.~\cite{Moncrief:1974am,Moncrief:1974an}. However, the tidal energy of the star that enters the binary equation of motion is only part of the total energy. We employ the calculation first discussed in~\cite{Feng:2024olt} to obtain the tidal energy of a compact star with a point mass orbiting in a circular orbit, and compare that with the prediction that was previously used assuming an effective tidal action in the Effective Field Theory approach and an effective mode description for the tidal action~\cite{Steinhoff:2016rfi}. The relative difference ranges from zero to $3\%$ if the inspiraling frequencies of a normal binary neutron star is $\leq 10^3\,{\rm Hz}$. Inaccuracy in the tidal energy may lead to a $\sim 0.5$ rad phase error in the gravitational waveform.

Therefore, the scattering formalism proves to be a powerful tool to describe the tidal response of stars under external gravitational drivings. With this formalism, both the transient mode energy and the dynamic tide energy can be computed accurately at the linear order of the external tidal gravitational field. This directly helps to improve the waveform accuracy in modeling matter effects in neutron star binaries. Combining with Post-Newtonian (PN) theory of nonlinear tides, a hybrid method may further reduce the model error. This will be particularly important for future observations with third-generation gravitational wave detectors, as the model systematics should be smaller than the detector noise to avoid biased measurements.

This paper is organized as follows. In Sec.~\ref{sec:scat} we introduce the wave-scattering formalism and discuss the duality between wave scattering and point-mass scenario, along with the definition of the external tidal field. In Sec.~\ref{sec:transient}, we provide a description of transient mode excitation in the relativistic regime. We derive the mode energy during resonant excitation and compare the results with the Newtonian case. Sec.~\ref{sec:sta} focuses on the steady excitation of the f-mode. In Sec.~\ref{sec:flux}, we establish a flux balance law that connects the total perturbative energy in spacetime to the scattering phase. We then calculate the tidal energy using the self-force approach in the point-mass scenario in Sec.~\ref{sec:eqt}, and incorporated it into the Effective-One-Body (EOB) framework in Sec.~\ref{sec:EOB}. The results of tidal energy and the gravitational-wave phase shifts are discussed in Sec.~\ref{sec:res}. We conclude in Sec.~\ref{sec:conclusion}.

\section{Scattering formalism}\label{sec:scat}

\begin{figure}
    %\centering
    \includegraphics[width=1.1\linewidth]{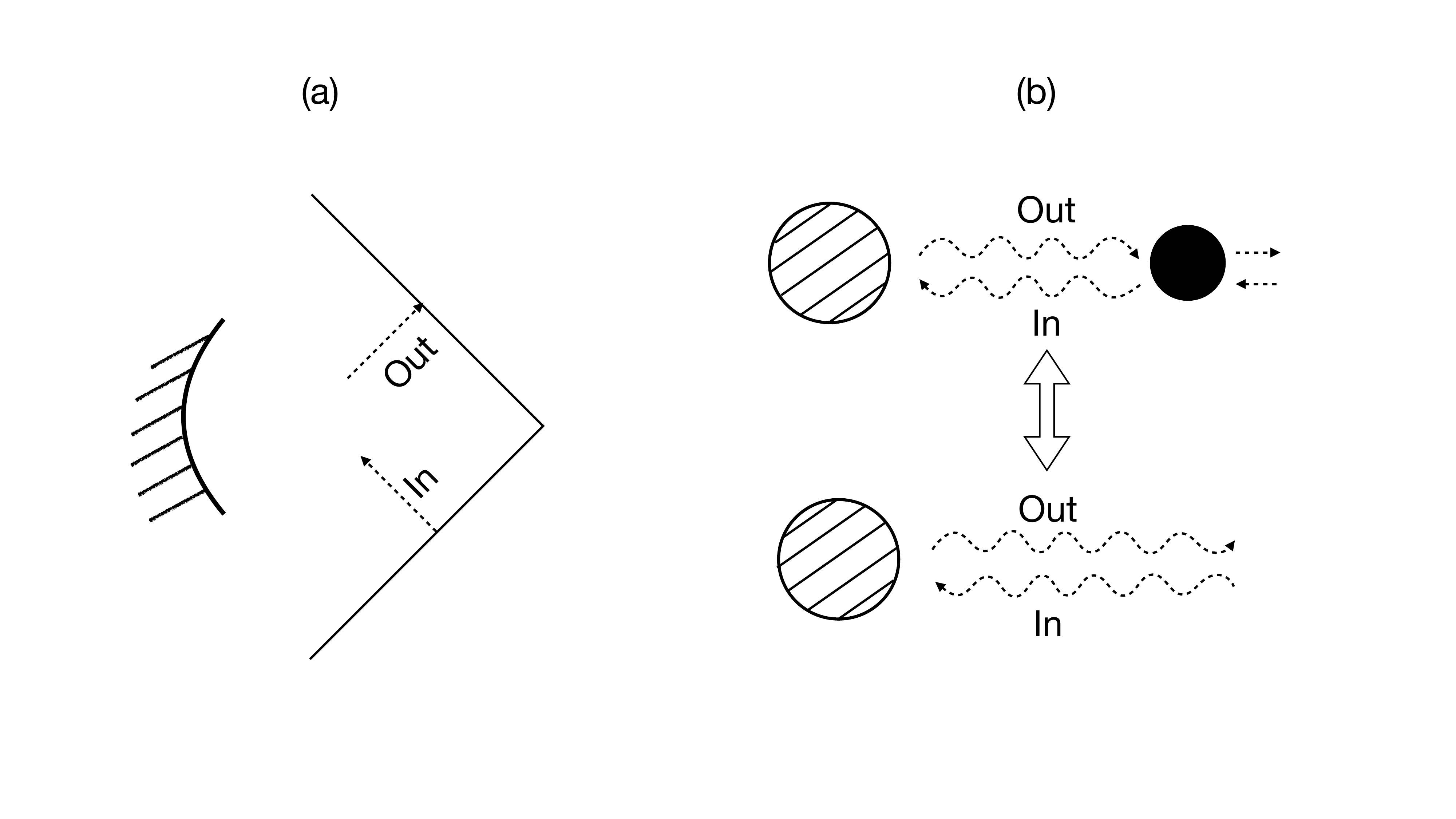}
    \caption{Schematic illustration of 
(a) Wave-scattering picture: gravitational waves are injected from past null infinity and scattered back to future null infinity by the star.
(b) Duality between the wave-scattering picture and point-mass scenario.}
    \label{fig:cartoon}
\end{figure}

We consider a star with mass $M$ and radius $R$. For simplicity, the star is assumed to be nonrotating, in which case the spacetime outside the star is Schwarzschild in the non-perturbed scenario. The gravitational perturbation in this vacuum exterior regime can be described by Regge-Wheeler-Zerilli formalism~\cite{Regge:1957td,Zerilli:1970se,Zerilli:1970wzz}. For a generic frequency component $\omega$ and in the limit that $r\to \infty$, the master variable $Z$ has the asymptotic behavior:
\begin{align}
Z(r\to \infty) = A_{\rm out}(\omega) e^{i \omega r_*}+A_{\rm in}(\omega) e^{- i \omega r_*}
\end{align}
where $r_*$ is the {{tortoise}} coordinate. With the $e^{-i \omega t}$ factor restored it is just
\begin{align}
Z(r\to \infty) e^{-i \omega t} = A_{\rm out}(\omega) e^{i \omega u}+A_{\rm in}(\omega) e^{ i \omega v}
\end{align}
with $u:= t-r_*$ and $v:=t+r_*$.

Notice the ingoing piece of the wave $A_{\rm in}$ represents the external gravitational driving, and the response of the star is encoded in the complex scattering/reflection coefficient (see Fig.~\ref{fig:cartoon})
\begin{align}
\mathcal{R} := \frac{A_{\rm out}}{A_{\rm in}}\,.
\end{align}

There are a few interesting properties of $\mathcal{R}$. First, for any real frequency $\omega$, due to energy conservation, the ingoing energy flux has to be equal to the outgoing energy flux at infinity, which means that $|\mathcal{R}|=1$. Second, the analytical structure of $\mathcal{R}$ should look like
\begin{align}\label{eq:rpole}
\mathcal{R} = \Pi_n \frac{\omega -\omega^*_n}{\omega-\omega_n} \times {\rm [nonpole \,\,component]}
\end{align}
with $\omega_n =\omega_{n {\rm R}} -i \gamma_n\,,\omega_{-n} =-\omega_{n {\rm R}} -i \gamma_n $ being the quasi-normal mode frequencies. The zeros in the numerator are paired with poles of $\mathcal{R}$ because under the time-reversal operation, we expect the role to switch between ingoing and outgoing waves, so that $\mathcal{R} \leftrightarrow 1/\mathcal{R}$.

The expression of Eq.~\eqref{eq:rpole} actually has non-trivial implications. Let us define the residue of $\mathcal{R}$ at the pole with frequency $\omega_n$ as
\begin{align}
{\rm Res}[\mathcal{R}(\omega)]:={\rm W}_n = \lim_{\omega \to \omega_n} (\omega-\omega_n)\mathcal{R}(\omega)\,.
\end{align}
Notice that another way to write down the residue is
\begin{align}
{\rm W}_n = \left . \frac{A_{\rm out}}{dA_{\rm in}/d \omega} \right |_{\omega_n} = 2 \omega_n \mathcal{B}_n\,,
\end{align}
where $\mathcal{B}_n$ is the {\it black hole excitation factor} in black hole perturbation theory~\cite{Leaver:1986gd}. It should be generalized to star modes in the scenario considered here.

On the other hand, in the case that  $\omega \approx \omega_n$, We may expand $\mathcal{R}$ as
\begin{align}\label{eq:Ranaly}
\mathcal{R} &=\frac{\omega -\omega^*_n}{\omega-\omega_n} \mathcal{R}_{0,n}(\omega) \nonumber \\
& = \left ( 1+\frac{2 i \gamma_n}{\omega-\omega_n} \right ) \mathcal{R}_{0,n}(\omega) \,.
\end{align}
Notice that $\mathcal{R}_{0,n}(\omega)$ no longer has a pole at $\omega_n$, so that
\begin{align}\label{eq:wnp}
{\rm W}_n=2 i \gamma_n \mathcal{R}_{0,n}(\omega_n)\,,
\end{align}
and 
\begin{align}\label{eq:wn}
|{\rm W}_n |=2  \gamma_n |\mathcal{R}_{0,n}(\omega_n)| \approx 2 \gamma_n\,,
\end{align}
or
\begin{align}
|\mathcal{B}_n| \approx \left |\frac{\gamma_n}{\omega_n}\right | \approx \frac{1}{Q_n}\,,
\end{align}
where $Q_n$ stands for the mode quality factor. The last two equations hold for modes with $\gamma_n \ll \omega_{n {\rm R}}$ and $\gamma_n \ll |\omega_n-\omega_m|$ for other modes with $m \neq n$. This ensures that $\mathcal{R}_{0,n}(\omega_n) \approx \mathcal{R}_{0,n}(\omega_{n {\rm R}})$ is true as $|\mathcal{R}_{0,n}(\omega)|=1$ is expected for real frequencies. Notice that the $\gamma_n \ll \omega_{nR}$ condition generally means that the amplitude of non-pole part of $\mathcal{R}_{0,n}$ is approximately one. If the condition $\gamma_n \ll |\omega_n-\omega_m|$ is not satisfied for certain mode $m$, we will need to take into the account the factor $|(\omega_n-\omega^*_m)/(\omega_n-\omega_m)|$ in the last equation of Eq.~\eqref{eq:wn}.
Nevertheless, when the conditions are satisfied, we observe that {\it the (half) amplitude of the residue of mode $n$ is essentially the decay rate of this mode, or the star excitation factor is the inverse of the mode quality factor}.
The physical meaning of residue ${\rm W}_n$ will become clear in the discussion of transient mode energy in Sec.~\ref{sec:transient}, and the property shown in Eq.~\eqref{eq:wn} will be useful to match the results with Newtonian scenarios.

\begin{figure}
    %\centering
    \includegraphics[width=0.9\linewidth]{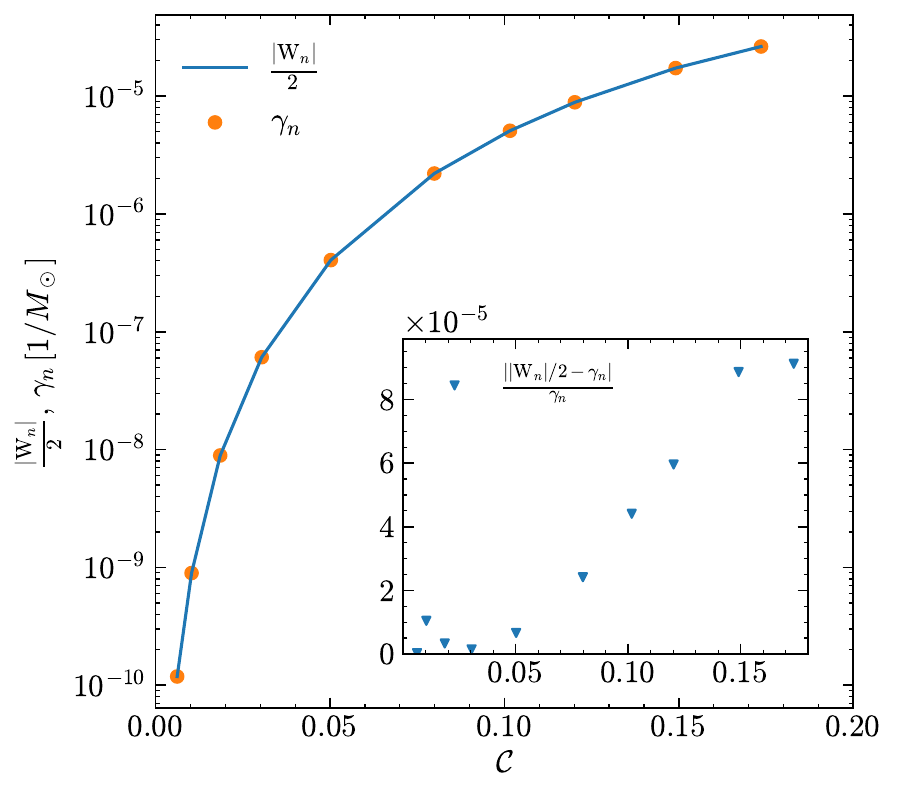}
    \caption{$|{\rm W}_n|$ and $\gamma_n$ as functions of stellar compactness, $\mathcal{C}:=M/R$, for f-mode. The results are calculated from a set of stellar models using the polytropic equation of state with $n=1$ and fixed stellar mass $M=1.4\,M_\odot$. The inset shows the relative difference between $|{\rm W}_n|$ and $\gamma_n$.}
    \label{fig:Wn-gamman}
\end{figure}

In order to verify this point, we construct a series of $n=1$ polytropic stars with the mass fixed to be $M=1.4 M_\odot$. Gravitational waves are injected from past null infinity and scattered by the star towards future null infinity. The details are presented in Appendices~\ref{sec:appendix pert} and \ref{sec:appendix residue}. 
The decay rate of a particular mode is obtained by identifying the poles on the complex frequency plane. On the other hand, ${\rm W}_n$ is measured by numerically evaluating the residue around the pole. For the $\ell=2$ f-mode, the comparison between $|{\rm W}_n|/2$ and $\gamma_n$ is presented in Fig.~\ref{fig:Wn-gamman} for the selected series of star configurations. It is evident that the relative difference is on the $10^{-4}$ level. 

As a side note, it turns out that this series-of-poles representation of the black hole excitation factor may also prove useful in understanding the spectral signatures of black holes. Consider a scenario where there are two poles $\omega_n, \omega_m$ near each other for certain range of black hole parameters. In this case, {we may write 
\begin{equation}
    \mathcal{R} = \frac{1}{\omega-\omega_n} \frac{1}{\omega-\omega_m}\mathcal{X}_{0,n,m}(\omega)\,,
\end{equation}
where $\mathcal{X}_{0,n,m}$ is a smooth function near $\omega_n$. We do not take the expansion like in Eq.~(\ref{eq:Ranaly}) because black holes are dissipative systems due to the event horizon. Nevertheless, we can still obtain
\begin{equation}
    {\rm W}_n = \frac{1}{\omega_n-\omega_m}\mathcal{X}_{0,n,m}(\omega_n)\,.
\end{equation}
}Since $|\omega_n-\omega_m| \ll |\omega_n|$, we find a large amplification factor in ${\rm W}_n$ and consequently the black hole excitation factor $\mathcal{B}_n$. This phenomenon has been observed numerically in the literature and is often referred to as the ``mode avoidance and resonant excitation effect ''~\cite{Onozawa:1996ux,Cook:2014cta,Berti:2003jh,Motohashi:2024fwt}. The argument here provides a natural explanation for this phenomenon. It is also straightforward to see that $\mathcal{B}_n \approx -\mathcal{B}_m$ for sufficiently close poles, so the excitation of these two modes cancel largely with each other.

The scattering formalism exhibits the convenient property of encoding the tidal response of the star into the input/output relation at infinity. It is also connected to the scenario with a point mass orbiting around the same star, as depicted in Fig.~\ref{fig:cartoon}. Inside the particle's orbital radius, the vacuum solution of the Zerilli function can still be written as
\begin{align}
Z(r) = A_{\rm out} Z_{\rm out}(r) + A_{\rm in} Z_{\rm in}(r)\,,
\end{align}
where $Z_{\rm out,in}$ are the homogeneous solutions that satisfies the $e^{\pm i \omega r_*}$ asymptotic boundary condition at infinity, {with $\omega$ related to the orbital frequency $\Omega$ through $\omega = m\Omega$}. 
{This implies that, within the wave-scattering picture, one can always fine-tune the amplitudes of the incoming and outgoing waves at infinity such that the solution inside the orbital radius matches the one obtained in the point-mass picture.}
Therefore, from the point of view of the star, whether it is driven by the gravitational field from a point mass or consistently tuned gravitational waves from infinity is indistinguishable from each other. There is always a wave-scattering problem that is dual to the point-mass scenario. We shall apply this view in studying the steady excitation of stars in Sec.~\ref{sec:sta}.

\subsection{Definition of the external tidal field}

In the effective field theory description of tidally perturbed bodies, the internal degrees of freedom of each body couples with external tidal fields. Although the definition of external fields is not unique, it is reasonable to define it as the field as if the target body were removed from the spacetime. In the scenario where a point mass orbits around a star as considered here, the 'external field'  of the star may be characterized by its 'equivalent' amplitude $A_{\rm in}$ at null infinity, according to the mapping relation shown in panel (b) of Fig.~\ref{fig:cartoon}. If the star were to be removed from spacetime, one can evaluate the value of the tidal tensor $\mathcal{E}_{ij}$ at $r=0$ assuming the same incoming wave amplitude at infinity. 
This defines an operational procedure to obtain the external tidal field for the model problem considered here, which allows us to compare the results measured both in $A_{\rm in}$ and in $\mathcal{E}_{ij}$.

As mentioned above, the ingoing wave acts as the external gravitational driving. 
Therefore, we can define the external tidal field in terms of the ingoing wave amplitude $A_{\rm in}$. 
Since this definition is independent of the nature of the central object, we consider the perturbations in Minkowski spacetime for simplicity. The perturbed metric in Regge-Wheeler gauge is given by \cite{Regge:1957td,Zerilli:1970se,Zerilli:1970wzz}
\begin{equation}
\begin{split}
    ds^2 = &-dt^2 + dr^2 + r^2(d\theta^2 + \sin^2\theta d\phi^2) \\
    &+ \Big[H_0 (dt^2+dr^2) + 2i\omega rH_1 dtdr \\
    &+ K r^2(d\theta^2 + \sin^2\theta d\phi^2)\Big]Y_{\ell m}(\theta, \phi) e^{-i\omega t } \,,
\end{split}
\end{equation}
and the Zerilli equation reduces to
\begin{equation}
    \frac{d^2Z}{dr^2} + \left[\omega^2-\frac{\ell(\ell+1)}{r^2} \right] Z=0\,,
\end{equation}
where 
\begin{align}
 K &= \frac{n+1}{r} Z + \frac{dZ}{dr}\,, \\ 
 H_1&= -\frac{Z}{r}-\frac{dZ}{dr}\,,\\
H_0 &=-\frac{1}{n } \left[ (\omega^2 r^2-n) K +  \omega^2 r^2 H_1 \right]\,,\label{eq:H0}\\
n &= \frac{(\ell-1)(\ell+2)}{2}\,.
\end{align}
The regular solution of Zerilli variable is $Z = A\omega r j_{\ell}(\omega r)$, which has the asymptotic behavior at spatial infinity, 
\begin{equation}
    Z(r\to \infty) = \frac{A}{2i}e^{i(\omega r-\ell\pi/2)}-\frac{A}{2i}e^{-i(\omega r-\ell\pi/2)}\,,
\end{equation}
from which the ingoing wave amplitude reads
\begin{equation}\label{eq:Ain}
    A_{\rm in} = -\frac{A}{2i}e^{i\ell\pi/2}\,.
\end{equation}
Substituting the solution of $Z$ into Eq.~\eqref{eq:H0} and taking the $r\to 0$ limit, we get 
\begin{equation}
\begin{split}
    H_0(r\to 0) &= A\omega \left[(n+2)j_\ell(\omega r) + r\frac{dj_\ell(\omega r)}{d r}\right] \\
    &= \frac{(\ell+1)(\ell+2)}{2(2\ell+1)!!}A\omega^{\ell+1}r^{\ell}+\mathcal{O}(r^{\ell+1})\,.
\end{split}
\end{equation}
The external tidal field can be decomposed as
\begin{equation}
    \mathcal{E}_{L} = \sum_{m=-\ell}^{\ell}\mathcal{E}_{\ell m}\mathcal{Y}_L^{\ell m}\,,
\end{equation}
where the symmetric traceless tensor $\mathcal{Y}_L^{\ell m}$ is defined
by 
\begin{equation}
    Y_{\ell m}(\theta, \phi) = \mathcal{Y}_L^{\ell m}n^L\,,
\end{equation}
with $\boldsymbol{n} = (\sin\theta\cos\phi, \sin\theta\sin\phi, \cos\theta)$. 
$L$ denotes a string of indices on a symmetric trace-free tensor, for example $n^{L=2}= n^in^j-\frac{1}{3}\delta^{ij}$. 
In our simplified case, the time-time metric component has the form~\cite{Thorne:1980ru}
\begin{equation}
    \frac{1+g_{tt}}{2} = \sum_\ell -\frac{1}{\ell !}r^\ell \mathcal{E}_L n^L\,,
\end{equation}
from which we can read the $(\ell,m)$ part of the tidal tensor through
\begin{equation}
    -\frac{1}{\ell!}r^\ell\mathcal{E}_{\ell m}Y_{\ell m} =  \frac{1}{2} H_0 Y_{\ell m}\,.
\end{equation}
The result is
\begin{equation}
    \mathcal{E}_{\ell m} = -\frac{(\ell+1)(\ell+2)\ell!}{4(2\ell+1)!!}A\omega^{\ell+1}\,,
\end{equation}
and 
\begin{equation}
    \mathcal{E}_{L}\mathcal{E}^{L}\big|_{\ell m} = \frac{(\ell+1)(\ell+2)(\ell+2)!}{16\pi(2\ell+1)!!}|A_{\rm in}|^2\omega^{2\ell+2}\,.
\end{equation}
For the $(\ell,m)=(2,2)$ mode, we have
\begin{equation}\label{eq:E22}
    \mathcal{E}_{ij}\mathcal{E}^{ij}\big|_{\ell m=22} = \frac{6}{5\pi} |A_{\rm in}|^2\omega^6\,.
\end{equation}

\begin{figure}
    %\centering
    \includegraphics[width=0.9\linewidth]{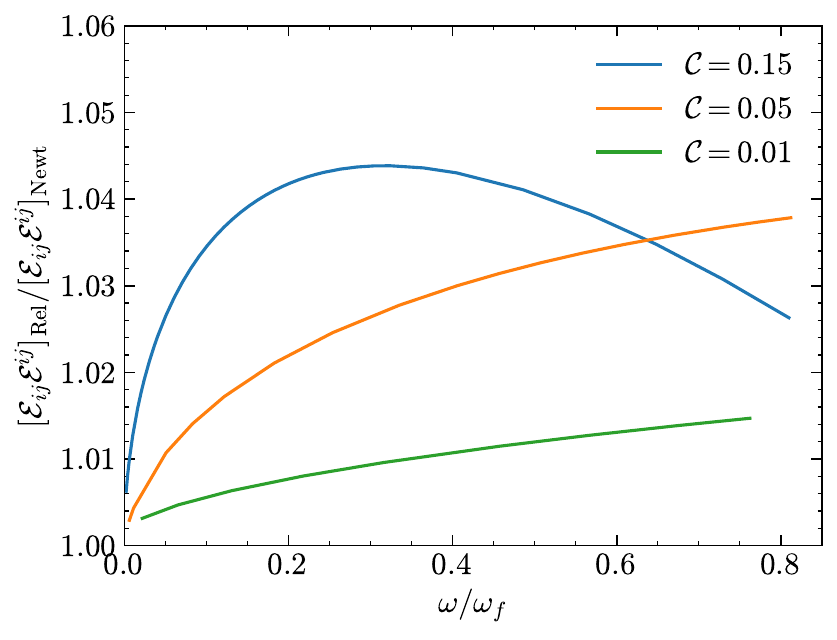}
    \caption{The ratio between the amplitude-based [Eq.~\eqref{eq:E22}] and Newtonian definitions ($[\mathcal{E}_{ij}\mathcal{E}^{ij}]_{\rm Newt}=\frac{9}{4}(\frac{m_2}{m_1+m_2})^2\Omega^4$) of external tidal field as function of the GW frequency $\omega=m\Omega$ for $(\ell,m)=(2,2)$ mode. $\omega_f$ is the f-mode frequency.}
    \label{fig:Eij}
\end{figure}

In Fig.~\ref{fig:Eij}, we compare the external tidal field defined via the ingoing amplitude extracted from perturbation calculations of the point-mass scenario (see Appendix~\ref{sec:pointmass}) with the Newtonian results. The difference between the amplitude-based definition and the Newtonian case is found to be at the $10^{-2}$ level.

\section{Transient excitation}\label{sec:transient}

In Newtonian theory, the excitation of modes of a star can be described via Eq.~\eqref{eq:newton}, which relies on the projection of external gravitational force into a mode basis. However, in the relativistic regime the wavefunction of an eigenmode of a star is generally defined in the entire spacetime domain, as the modes radiate. In addition, modes do not form a complete basis for relativistic stars. As a result, it is not necessarily clear how to promote the inner product in Eq.~\eqref{eq:newton} to the relativistic regime. Note that for black holes, a bilinear form can be defined to ensure mode orthogonality behavior~\cite{Ma:2024qcv}. It is still unknown whether or how a similar construction may be achieved for stars.

In this section, we adopt an alternative approach. We show that the transient mode excitation can already be described with the input/output relation in the wave-scattering scenario. We compute the transient mode energy as an evolving binary sweeps through the frequency of a resonant mode in Sec.~\ref{sec:tr1}, and show that it is indeed consistent with the known Newtonian formula in the Newtonian regime in Sec.~\ref{sec:tr2}.
The difference between the Newtonian description and the relativistic description is discussed in Sec.~\ref{sec:tr3}.

\subsection{Deriving the transient mode excitation}\label{sec:tr1}

In order to model a time-dependent driving field, we promote the ingoing gravitational wave in the scattering formalism to be time-dependent.
Let us consider the case that the driving frequency slowly shifts in time, $\dot \sigma \ll \sigma^2$ {(In this subsection, we denote the driving frequency by $\sigma$ to avoid confusion with the Fourier frequency $\omega$.)}:
\begin{align}
A_{\rm in}(t) =A_{\rm in,0}e^{-i \int^t dt' \sigma(t')}\,,
\end{align}
so that
\begin{align}\label{eq:aino}
A_{\rm in}(\omega) & = A_{\rm in,0}\int dt_1 e^{i \omega t_1} e^{-i \int^{t_1}_0 dt' \sigma(t')} \nonumber \\
&= A_{\rm in,0}\int dt_1  e^{i \int^{t_1}_0 dt' (\omega- \sigma(t'))}\,.
\end{align}

We define a characteristic time $t_\omega$ that is associated to the driving frequency $\sigma$  via $\sigma(t_\omega) =\omega$. For $t$ close to $t_\omega$, $\sigma(t)$ can be expanded as $\sigma(t) =\omega +\dot \sigma \Delta t$ ($\Delta t :=t-t_\omega$). Plugging in these simplification into Eq.~\eqref{eq:aino} we get
\begin{align}
A_{\rm in}(\omega) &= A_{\rm in,0} e^{i \Phi_\omega} \int d \Delta t e^{i \dot{\omega} \Delta t^2/2} \nonumber \\
&= A_{\rm in,0} e^{i \Phi_\omega} (1+i) \sqrt{\frac{\pi}{\dot \sigma}}\,,
\end{align}
where $\Phi_\omega = \int^{t_\omega}_0 dt' (\omega- \sigma(t'))$.
In the frequency domain we still have the input-output relation with
\begin{align}
A_{\rm out}(\omega) = \mathcal{R}(\omega) A_{\rm in}(\omega)\,,
\end{align}
where $\mathcal{R}(\omega)$ is essentially the Fourier transform of the transfer function. The outgoing wave in the time domain is given by
\begin{align}
A_{\rm out} =\frac{1}{2\pi}\int^\infty_{-\infty} d\omega e^{-i \omega t }\mathcal{R}(\omega) A_{\rm in}(\omega)\,.
\end{align}

For sufficiently late $t$ we close the contour on the lower half-plane, and the poles of $\mathcal{R}(\omega)$ give rise to the quasi-normal mode part of $A_{\rm out}$, where the direct part and tail part are not the focus here. We have

\begin{align}
&A_{\rm out, mode} \nonumber \\
&=\frac{1+i}{2\pi} \sqrt{\frac{\pi}{\dot \sigma}} A_{\rm in,0}  \sum_n e^{-i \int^{t_{\omega_n}} dt' \sigma(t')} 2 \pi i {\rm W}_n e^{-i \omega_n (t-t_{\omega_n})} \nonumber \\
&\approx \frac{1+i}{2\pi} \sqrt{\frac{\pi}{\dot \sigma}} A_{\rm in,0}  \sum_n  e^{-i \int^{t_{\omega_{n {\rm R}}}} dt' \sigma(t')} 2 \pi i {\rm W}_n e^{-i \omega_n (t-t_{\omega_n})} \nonumber \\
& =\frac{1+i}{2\pi} \sqrt{\frac{\pi}{\dot \sigma}} \sum_n  A_{\rm in}(t_{\omega_{n {\rm R}}}) 2 \pi i {\rm W}_n e^{-i \omega_n (t-t_{\omega_n})}\,,
\end{align}
for modes sufficiently close to the real axis. In this way we can find out the amplitude of freely excited modes after transient resonances.

This can already be used to obtain the transient energy of a single mode. After a certain mode is resonantly excited, the presence of a mode piece $A_{\rm out,mode,n}$ contributes additional flux in the outgoing radiation. This free-mode piece (with frequency $\omega_{n{\rm R}}-i \gamma_n$) should quickly dephase with the evolving orbital frequency $\sigma$. Physically this corresponds to a free mode of star excited in addition to the driven oscillation. Notice that the energy of a free mode can be computed from its corresponding integrated flux at infinity:
\begin{align}\label{eq:mode energy}
E_{\rm mode,\,Rel} &= C_{\omega_{n{\rm R}}} \int^\infty_{t_{\omega_n}} dt|A_{\rm out, mode,n}|^2 \nonumber \\
& = C_{\omega_{n{\rm R}}} \frac{2 \pi |{\rm W}_n|}{\dot{\sigma}} |A_{\rm in}(t_{\omega_{n{\rm R}}})|^2\nonumber \\
& = \frac{1}{32}\frac{(\ell+2)!}{(\ell-2)!}\frac{|{\rm W}_n|}{\dot\sigma}|A_{\rm in}(t_{\omega_{n{\rm R}}})|^2\omega_{n{\rm R}}^2\,,
\end{align}
where we have used Eq.~\eqref{eq:wn} and the geometric factor $C_\omega$ is given by~\cite{Zerilli:1970wzz}
\begin{align}
C_{\omega} = \frac{1}{64\pi}\frac{(\ell+2)!}{(\ell-2)!}\omega^2\,.
\end{align}

In fact, the problem can also be understood from an alternative perspective. The resonance occurs over a finite time interval of duration $\Delta t \sim \sqrt{1/\dot{\sigma}}$, centered around the resonance point ($\sigma(t_{\omega_{n\rm R}})=\omega_{n{\rm R}}$). During this period, energy transfer takes place due to the beating between the driven and free components, which is imprinted as an interference pattern in the outgoing radiation. The energy deposited into the mode during resonance can be directly computed from
\begin{equation}\label{eq:mode energy2}
\begin{split}
    E_{\rm mode,\,Rel} &= \int_{t_1}^{t_2} C_\sigma \left( |A_{\rm in}(\sigma)|^2 - |A_{\rm out}(\sigma)|^2 \right) dt\\ 
    &= \int_{\sigma_1}^{\sigma_2} C_\sigma |A_{\rm in}(\sigma)|^2 \left[ 1 - |R(\sigma)|^2 \right] \frac{1}{\dot{\sigma}} d\sigma\\
    &= \int_{\sigma_1}^{\sigma_2}C_\sigma|A_{\rm in}(\sigma)|^2\frac{4\gamma_n^2}{(\sigma-\omega_{n{\rm R}})^2+\gamma_n^2}\frac{1}{\dot\sigma}d\sigma\\
    &\approx C_{\omega_{n\rm R}}|A_{\rm in}(\omega_{n\rm R})|^2\frac{1}{\dot\sigma}\int_{-\infty}^{\infty}\frac{4\gamma_n^2}{(\sigma-\omega_{n{\rm R}})^2+\gamma_n^2}d\sigma\\
    & = C_{\omega_{n\rm R}}|A_{\rm in}(\omega_{n\rm R})|^2\frac{4\pi\gamma_n}{\dot\sigma}\\
    &=\frac{1}{16}\frac{(\ell +2)!}{(\ell-2)!}\frac{\gamma_n}{\dot \sigma}|A_{\rm in}(\omega_{n\rm R})|^2\omega_{n\rm R}^2\,.
\end{split}
\end{equation}
In the third line, we have used the analytic structure of $\mathcal{R}$ near the mode pole, as given in Eq.~\eqref{eq:Ranaly}, which implies that $1 - |\mathcal{R}(\sigma)|^2$ takes a Lorentzian form centered at $\sigma = \omega_{n\rm R}$ with width $\gamma_n$. In the fourth line, we extended the integration limits to $(-\infty, \infty)$, justified by the fact that the contribution of the integral comes from frequency interval $\sim \gamma_n$, which is much less than the resonance width, i.e., $\sigma_2-\sigma_1\sim \sqrt{\dot \sigma}\gg \gamma_n $. Meanwhile, the integral can be approximated by its value at the resonance point, i.e., $C_\sigma \approx C_{\omega_{n\rm R}}$ and $|A_{\rm in}(\sigma)|^2 \approx |A_{\rm in}(\omega_{n\rm R})|^2$.
It is obvious Eq.~\eqref{eq:mode energy2} agree with Eq.~\eqref{eq:mode energy}. This calculation also confirms the understanding that external driving deposits the resonant mode energy $ E_{\rm mode,\,Rel}$ to the star during the transient resonance period, which is later on carried away by gravitational wave radiation.

The transient energy of the modes is directly proportional to $|{\rm W}_n|$, which is proportional to the damping rate of the mode $\gamma_n$.
For a normal neutron star with mass $\sim 1.4\,M_\odot$ and radius $\sim 13\,{\rm km}$, the ratio between $\gamma_n$ and $\omega_{nR}$ is $\sim 10^{-3}$ for a $\ell=2$ f-mode and $10^{-14}$ for a $\ell=2$ g$_1$-mode. 
Therefore, it is computationally challenging to calculate $|{\rm W}_n|$ for various modes. In Appendix~\ref{sec:appendix residue} and Appendix~\ref{sec:alter}, we illustrate two independent approaches for computing $|{\rm W}_n|$: the method discussed in Appendix~\ref{sec:appendix residue} is based on a direct evaluation of the residue around the pole, and the method discussed in Appendix~\ref{sec:alter} is motivated by the eigenvalue perturbation theory~\cite{Yang:2014tla,Ma:2024qcv}.

As Eq.~\eqref{eq:mode energy} generally applies for various kinds of mode excitation during a binary inspiral process, it is instructive to compute $|{\rm W}_n|$ of various relevant modes discussed in the literature. This systematic mode characterization will be studied in a separate work.

\subsection{Newtonian limit}\label{sec:tr2}

In the Newtonian regime, we may reexpress Eq.~\eqref{eq:newton} as
\begin{equation}\label{eq:newton an}
    \ddot a_n +\omega_n^2 a_n = -\frac{\mathcal{E}_{\ell m}Q_{n\ell m}}{\ell!}\,,
\end{equation}
where $a_n$ is the mode amplitude. The tidal overlap integral is defined as usual~\cite{Lai:1993di},
\begin{equation}
    Q_{n\ell m} = \int \ell\rho r^{\ell+1}[
\xi^r_n+(\ell+1)\xi^\perp_n)]dr\,,
\end{equation}
with $\boldsymbol{\xi}_n = (\xi^r_n \boldsymbol{e}_r+ \xi^\perp_n r\boldsymbol{\nabla}_\perp) Y_{\ell m}$ is the mode wavefunction. After the resonance, we expect $a_n$ to oscillate at its own eigenfrequency $\omega_n$, thus it is more appropriate to define a function $c_n$ such that $a_n  = c_ne^{-i \omega_nt}$. Plugging $c_n$ into Eq.~\eqref{eq:newton an} and dropping the $\ddot c_n$ term, we get
\begin{equation}
    c_{n} = \frac{1}{2i\omega_n} \frac{\mathcal{E}_{\ell m}Q_{n\ell m}}{\ell!}\int e^{i\dot\omega t^2/2}dt =  \frac{1+i}{2i\omega_n} \frac{\mathcal{E}_{\ell m}Q_{n\ell m}}{\ell!}\left(\frac{\pi}{\dot\omega}\right)^{1/2}\,,
\end{equation}
where we have used the stationary phase approximation. 
The resonant mode energy is then obtained with
\begin{equation}\label{eq:Newt energy}
\begin{split}
    E_{\rm mode,\,Newt} =& \omega_n^2|a_n|^2\\
     = & \frac{\pi}{2\dot \omega} Q_{n\ell m}^2 \frac{\mathcal{E}_{\ell m}^2}{(\ell!)^2}\\
    = &\frac{\pi}{2\dot \omega} Q_{n\ell m}^2 \left[\frac{(\ell+2)(\ell+1)}{2(2\ell+1)!!}\right]^2 |A_{\rm in}|^2\omega_n^{2\ell+2}\ .
\end{split}
\end{equation}

\begin{figure}
    %\centering
    \includegraphics[width=0.9\linewidth]{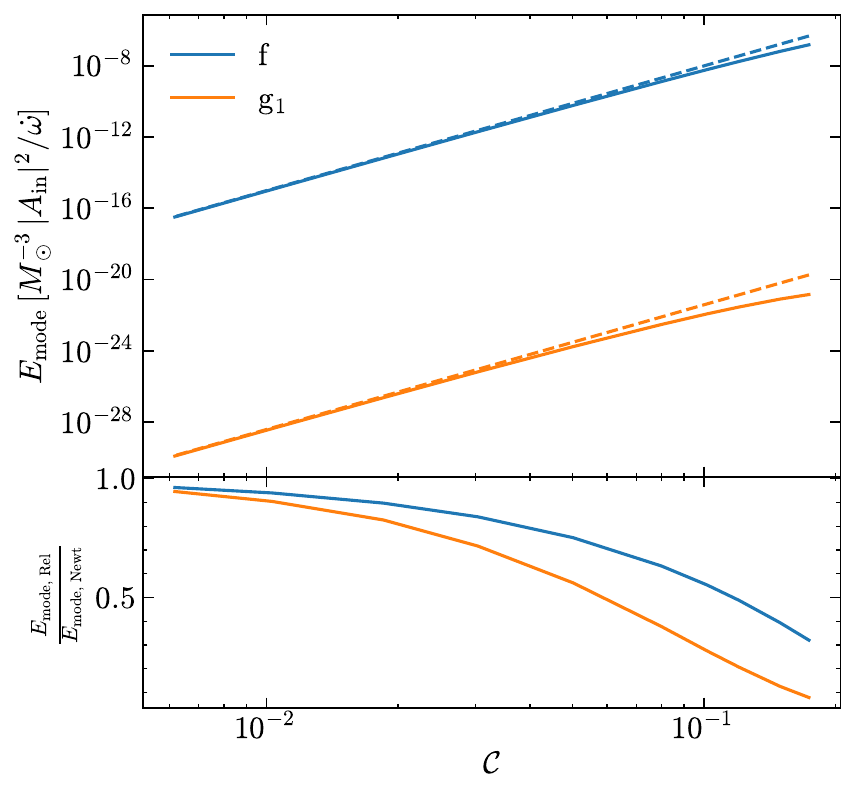}
    \caption{Top: Transient mode energy of f-mode (blue) and g$_1$-mode (orange) resonance as a function of the stellar compactness $\mathcal{C}$. Both results calculated from the relativistic formula [Eq.~\eqref{eq:mode energy}, solid] and the Newtonian formula [Eq.~\eqref{eq:Newt energy}, dashed] are plotted. The stellar models are constructed from polytropic equation of state with $n=1$. Bottom: Comparison of the transient energy between the two formula.}
    \label{fig:transient}
\end{figure}

\begin{figure}
    %\centering
    \includegraphics[width=0.9\linewidth]{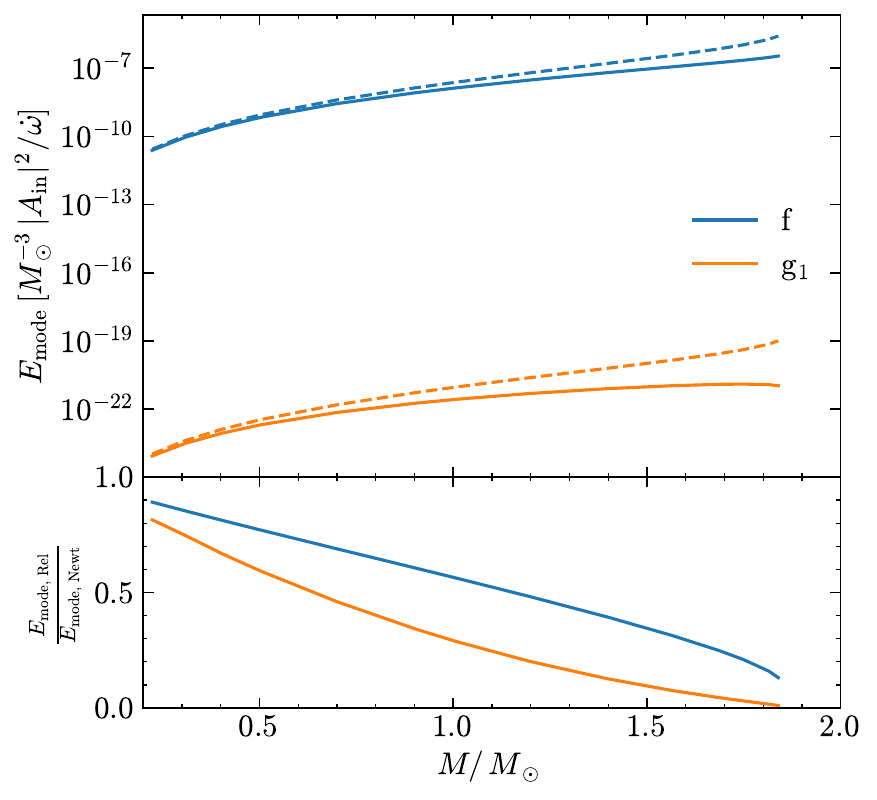}
    \caption{Same as Fig.~\ref{fig:transient}, but plotted as a function of stellar mass $M$. In this figure, the relativistic stars are modeled using a polytropic equation of state with $n=1$ and $\kappa = 86\,M_\odot^{2}$. The corresponding Newtonian stars are constructed to have the same masses and radii as the relativistic models.}
    \label{fig:transient_M}
\end{figure}

On the other hand, if we consider a damped mode with the amplitude $a_n = c_n e^{-i\omega_n t-\gamma_n t}$, the mode energy is 
\begin{equation}\label{eq:damped mode energy}
    E_{\rm mode,\,Newt} = \omega_n^2|c_n|^2e^{-2\gamma_n t}\,.
\end{equation}
The mode damping due to GW radiation can also be approximated using the multipole formula~\cite{Thorne:1980ru}, 
\begin{equation}\label{eq:damping}
\begin{split}
    \frac{dE_{\rm mode,\,Newt}}{dt} = & -\frac{(\ell+1)(\ell+2)}{\ell(\ell-1)\ell!(2\ell+1)!!} \, \langle {I}_L^{(\ell+1)} {I}_L^{(\ell+1)} \rangle \\
    =&-\frac{4\pi(\ell+1)(\ell+2)}{\ell(\ell-1)[(2\ell+1)!!]^2} Q_{n\ell m}^2\omega_n^{2\ell+2}|a_n|^2\\
    =& -\frac{4\pi(\ell+1)(\ell+2)}{\ell(\ell-1)[(2\ell+1)!!]^2} Q_{n\ell m}^2\omega_n^{2\ell+2} |c_n|^2e^{-2\gamma_n t}\,,
\end{split}
\end{equation}
Combining Eq.~\eqref{eq:damping} with Eq.~\eqref{eq:damped mode energy},  we deduce the decay rate to be 
\begin{equation}\label{eq:Imomega}
    \gamma_n = \frac{2\pi}{[(2\ell+1)!!]^2}\frac{(\ell+1)(\ell+2)}{\ell(\ell-1)}Q_{n\ell m}^2\omega_{n}^{2\ell}\,.
\end{equation}
Plugging Eq.~\eqref{eq:Imomega} into Eq.~\eqref{eq:Newt energy}, we find
\begin{equation}\label{eq:mode energy newt}
    E_{\rm mode,\,Newt} = \frac{1}{16}\frac{(\ell+2)!}{(\ell-2)!}\frac{\gamma_n}{\dot\omega}|A_{\rm in}|^2\omega_{n}^2\,.
\end{equation}

In Sec.~\ref{sec:tr1}, we have computed the energy gained by stellar modes during resonant excitation in the fully relativistic regime, $E_{\rm mode,\,Rel}$, as given by Eq.\eqref{eq:mode energy} [or Eq.\eqref{eq:mode energy2}]. We expect that in the Newtonian limit these expressions should reduce to the energy derived in the Newtonian framework presented in this subsection $E_{\rm mode,\,Newt}$. A direct comparison between Eq.\eqref{eq:mode energy} (or Eq.\eqref{eq:mode energy2}) and Eq.\eqref{eq:mode energy newt} clearly shows that the Newtonian limit is correctly recovered (see also Fig.~\ref{fig:transient}). In the next subsection, we will numerically compare $E_{\rm mode,\,Rel}$ and $E_{\rm mode,\,Newt}$ for stars with different compactness, highlighting the deviations between the two formulas.

\subsection{Comparison with Newtonian stars for f-mode and g-mode resonances}\label{sec:tr3}
In this subsection, we present illustrative comparisons of the transient mode energy associated with f-mode and g-mode resonances between relativistic and Newtonian stars.
Notice that the f-mode is unlikely to be fully excited in a realistic binary.

We adopt a set of energy-polytrope equations of state to construct the stellar model~\footnote{For Newtonian star, we take the usual polytropic equation of state, $p=\kappa \rho^{1+1/n}$, where $\rho$ is the rest mass density.}, i.e., 
\begin{equation}
    p= \kappa \varepsilon^{1+1/n}\,,
\end{equation}
where $p$ is the pressure and $\varepsilon$ is the energy density. 
We set $n=1$ in this work for simplicity. 
We then fix the stellar mass to $M=1.4\,M_\odot$, and vary the parameter $\kappa$ to adjust the stellar radius $R$, thus producing stars with different compactness $\mathcal{C}:=M/R$. 
In this way, we construct a set of relativistic stars and a corresponding set of Newtonian stars, each spanning a range of stellar compactness.
To model the stratification relevant to g-modes, we also consider a parameterized setup with $\Gamma=\gamma(1+\delta)$, where for relativistic stars
\begin{align}
    \gamma :=& \frac{p+\varepsilon}{p}\frac{dp}{d\varepsilon}\,,\\
    \Gamma :=&\frac{p+\varepsilon}{p}\left(\frac{\partial p}{\partial \varepsilon}\right)_s\,.
\end{align}
Here, the subscript ``s'' indicates that the partial derivative is taken at constant composition (i.e., no change in entropy or chemical makeup).
Physically, $\gamma$ is the adiabatic index associated with the equilibrium structure, while $\Gamma$ governs the response of the stellar fluid to pulsational compressions with no composition change.
The difference $\Gamma - \gamma$ quantifies the degree of stable stratification in the stellar interior, which are essential for the existence and behavior of g-modes. {We take $\delta = 0.005$ in our calculation}~{\cite{Xu:2017hqo,Kuan:2021jmk}}.

The mode calculations for relativistic stars are provided in Appendices~\ref{sec:appendix pert} \& \ref{sec:appendix residue}, while the mode calculations for Newtonian stars can be found elsewhere (e.g., Ref.~\cite{Lai:1993di}). 
Here, we only compute the f-mode and g$_1$-mode for $\ell=2$.
We then use Eq.~\eqref{eq:mode energy} to calculate the transient mode energy for relativistic stars, while use Eq.~\eqref{eq:Newt energy} to calculate it for Newtonian stars.
The results are shown in Fig.~\ref{fig:transient} and Fig.~\ref{fig:transient_M}. In Fig.~\ref{fig:transient}, we plot the mode energy as a function of stellar compactness by fixing the mass at $M=1.4\,M_\odot$, keeping the same equation of state and varying the radius. 
In Fig.~\ref{fig:transient_M}, we instead plot the mode energy as a function of stellar mass, where the relativistic stars are constructed using a polytropic equation of state with index $n=1$ and polytropic constant $\kappa = 86\,M_\odot^2$, and the Newtonian stars are correspondingly set to have identical masses and radii as their relativistic counterparts~\footnote{As noted in Ref.~\cite{Lai:1993di}, the overlap integral $Q_{n\ell m}$—or equivalently, the relativistic quantity ${\rm W}_n$ used in our work—is highly sensitive to the mode frequency. To enable a meaningful comparison between the Newtonian and relativistic cases, we therefore aim to keep the mode frequencies as close as possible in the comparison. Specifically, we set the Newtonian stars to have the same radii as their relativistic counterparts, which ensures that the mode frequencies differ by less than 10\%. This choice  ensures that any differences in the resulting mode energies primarily reflect genuine physical differences rather than trivial frequency mismatches. We caution that this comparison is inherently \textbf{phenomenological}, as a fully consistent matching between Newtonian and relativistic stellar models is impossible.}.
It is seen that, as the compactness approaches zero (the Newtonian limit), the energy computed from the relativistic formalism agrees well with the Newtonian result. 
However, as the stellar compactness increases, the transient energy predicted by the relativistic formalism becomes significantly lower than that of the Newtonian case. For typical neutron star compactness ($\mathcal{C}\sim 0.15$), the energy associated with the resonances in the f-mode and g$_1$-mode is reduced by approximately $\sim 60\%$ and $\sim 90\%$, respectively, compared to the Newtonian case.
This means that a Newtonian calculation for the mode energy may receive a systematic error of up to one order of magnitude. As ${\rm W}_n$ is proportional to the square of the ``overlap integral” between the external tidal field and mode wavefunction, it suggests that similar overlap functions, such as those used for computing nonlinear mode-coupling coefficients, may also receive significant relativistic correction. In order to accurately describe the resonant mode excitations, it is necessary to adopt the description in the relativistic regime.

\section{Steady excitation}\label{sec:sta}
In this section, we discuss how steady excitation of modes can be described using the input-output relation within the wave scattering framework. We begin by establishing a flux balance law which connects the total energy stored in the system to the scattering phase. We numerically verify this formula using the Hamiltonian of a compact star derived in~\cite{Moncrief:1974am,Moncrief:1974an}. In the second part, we discuss an alternative way of computing the tidal energy of a binary containing compact star(s), motivated by gravitational self-force, and compare the predictions with those from an effective field description, which has been widely used in the previous literature.  

\subsection{Flux balance law}\label{sec:flux}

Considering a star driven by a steady strain of ingoing gravitational waves as depicted in Fig.~\ref{fig:cartoon}, one important question is what the total energy (associated with the driving field) within the spacetime should be. Notice that for characterizing the tidal effects within the inspiral waveform, currently the tidal energy of a deformed neutron star within a binary is computed by an effective field theory approach (see, e.g.~\cite{Steinhoff:2016rfi}), where all the degrees of freedom are summarized into one single mode. It is necessary to understand the systematic error associated with this effective one-mode approach by a direct comparison with a relativistic calculation.

According to the discussion in~\cite{Moncrief:1974am,Moncrief:1974an}, the energy associated with perturbations in a Schwarzschild spacetime and the energy associated with (fluid and gravitational) perturbations of a compact star can be derived in a Hamiltonian formalism. Alternatively, we shall show that the information about the energy of the star is also encoded in the phase of the scattered wave, which can be found from the input-output relation. Two descriptions actually lead to the same result, as demonstrated from numerical examples. 

Let us start with the analysis within the input-output formalism. In order to compute the energy at the steady state, it is instructive to consider a ``ramping-up'' stage where the field increases from zero to the steady value, and the energy of the spacetime should be equal to the net flux injected from infinity. 
This can be achieved by assuming a small change in driving frequency, $\omega\to\omega+i\epsilon$, such that the time dependence of the waves becomes $e^{- i \omega t} e^{\epsilon t}$, corresponding to a slowly growing amplitude. In this case, the outgoing amplitude takes the form
\begin{equation}
    A_{\rm out}  = A_{\rm in}\mathcal{R}(\omega+i\epsilon) = A_{\rm in} \left[\mathcal{R}(\omega)+\frac{\partial \mathcal{R}}{\partial \omega}i\epsilon\right]\,.
\end{equation}
Imagine $A_{\rm in,0}$ is an initial ingoing amplitude and it takes a long time $t$ to grow to the current amplitude  $A_{\rm in, t}$:
\begin{equation}
A_{\rm in, t} =e^{\epsilon t} A_{\rm in, 0}\,. 
\end{equation}
Then the total energy injected into the system should be equal to the accumulated differential flux over time, i.e.,
\begin{equation}
\begin{split}
    E_{\rm tot} & = C_\omega \int^t dt (|A_{\rm in,t}|^2-|A_{\rm out,t}|^2) \nonumber \\
    & =C_\omega\int^t dt |A_{\rm in,t}|^2 (1-|\mathcal{R}(\omega+i \epsilon)|^2) \nonumber \\
    & = -iC_\omega \int^t \epsilon dt |A_{\rm in,0}|^2 e^{2 \epsilon t} \left (\mathcal{R}^* \frac{\partial \mathcal{R}}{\partial \omega} -\mathcal{R}\frac{\partial \mathcal{R}^*}{\partial \omega}\right ) \nonumber \\
    & = -i \frac{C_\omega}{2} \left (\mathcal{R}^* \frac{\partial \mathcal{R}}{\partial \omega} -\mathcal{R}\frac{\partial \mathcal{R}^*}{\partial \omega}\right )|A_{\rm in,t}|^2\,.
\end{split}
\end{equation}
Reexpressing $\mathcal{R}(\omega)$ as $|\mathcal{R}(\omega)|e^{i \theta}$ ($|\mathcal{R}(\omega)| =1$ for real $\omega$), we then obtain
\begin{equation}\label{eq:total energy}
    E_{\rm tot} =  C_\omega \frac{\partial \theta}{\partial \omega } |A_{\rm in,t}|^2\,,
\end{equation}
from which we see that the total energy is proportional to the derivative of the scattering phase. 
It should be note that in Eq.~\eqref{eq:total energy}, the definition of total energy includes an arbitrary zero-point freedom. Physically it corresponds to the ``free-propagation'' time or phase that is not associated with the tidal response of the central object. In the analysis in~\cite{Ivanov:2022qqt} for the tidal response of black holes, a far-zone phase $\theta_{\rm far}$ is removed due to similar reasons. To remove this degree of freedom, we may consider the energy difference between two systems (I and II) that have the same external spacetime if there is no external perturbations, because the far-zone phase should be the same for these two systems.
Since in this work we are interested in the tidal effects, it is convenient to set the central object in system II as a black hole (with the same mass as the compact star), while keeping the compact star in system I. Then the energy reads
\begin{equation}\label{eq:e diff NS-BH}
    E_{\rm tot} = C_\omega \frac{\partial (\theta_{\rm CS}-\theta_{\rm BH})}{\partial\omega}|A_{\rm in}|^2\,.
\end{equation}
Note we have omitted the subscript ``t'' in $A_{\rm in}$.

It is important to emphasize that the energy here refers to the total energy difference, which includes not only the internal kinetic and potential energy of the compact star (similar to the term $\dot Q_L\dot Q^L+\omega_n^2Q_LQ^L$ in its Newtonian counterpart), as well as the coupling energy between the multipole moments of the compact star and the tidal field (similar to the term $Q_LE^L$ in its Newtonian counterpart), but also the energy of the perturbed spacetime (GW) itself. 

\begin{figure}
    %\centering
    \includegraphics[width=0.9\linewidth]{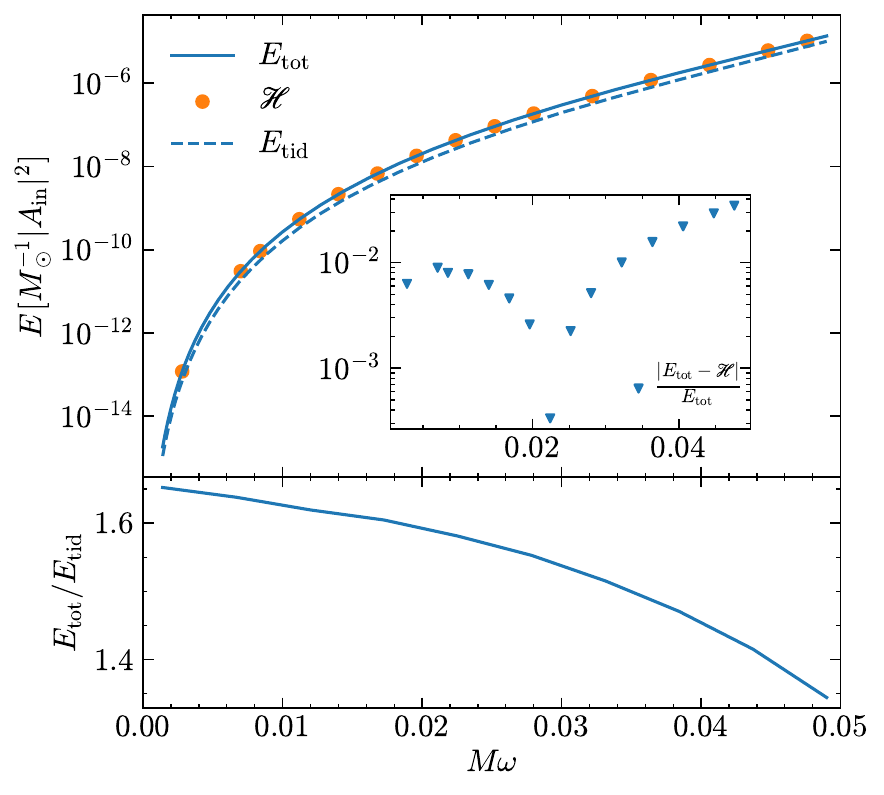}
    \caption{Comparison between the total energy obtained from the input-output formalism ($E_{\rm tot}$) and the energy derived from the Hamiltonian formalism ($\mathscr{H}$). The Compact star/Black hole mass is fixed at $M=1.4\,M_\odot$. Also shown is the tidal energy $E_{\rm tid}$. The inset shows the relative difference between $E_{\rm tot}$ and $\mathscr{H}$.}
    \label{fig:flux balance}
\end{figure}

We also perform a consistency check of the energy expression Eq.~\eqref{eq:e diff NS-BH} by comparing it with the Hamiltonian of the perturbation derived in Refs.~\cite{Moncrief:1974am,Moncrief:1974an}. The details of the Hamiltonian formalism are presented in Appendix~\ref{sec: appendix H}.
In Fig.~\ref{fig:flux balance} we present a representative example, where both the compact star and the black hole are set to $M=1.4\,M_\odot$, and the compactness of the compact star is chosen to be $\mathcal{C}=0.15$. We compare the total energy $E_{\rm tot}$, calculated by the flux balance method [Eq.~\eqref{eq:e diff NS-BH}], with the Hamiltonian difference $\mathscr{H}=\mathscr{H}_{\rm CS}-\mathscr{H}_{\rm BH}$, obtained from the Hamiltonian formalism. The excellent agreement between the two confirms the consistency of the two methods.

In this subsection, we have derived the total energy $E_{\rm tot}$ of the spacetime under steady driving, established its connection to the scattering coefficient (i.e., $E_{\rm tot}\propto \partial\theta/\partial\omega$), and confirmed the result numerically using the Hamiltonian formalism. Although this discussion is of physical interest and provides valuable insight into the wave scattering picture, $E_{\rm tot}$ is not the quantity we ultimately seek--namely, the star's tidal binding energy $E_{\rm tid}$. As discussed above, $E_{\rm tot}$ includes not only the star’s tidal energy, but also the energy carried by the driving field. This distinction is clearly illustrated in Fig.~\ref{fig:flux balance}, where we also compare $E_{\rm tot}$ with the tidal energy $E_{\rm tid}$ calculated in the next subsection. As expected, $E_{\rm tot} \neq E_{\rm tid}$.

Therefore, in Sec.~\ref{sec:eqt}, we will employ an self-force approach to compute the tidal energy of the star in the extreme mass ratio limit. We then further rescale this result to comparable-mass systems in Sec.~\ref{sec:EOB}, using the EOB framework.

\subsection{Equilibrium tidal energy in the extreme mass ratio limit}\label{sec:eqt}

We now focus on the computation of the tidal energy, which enters the binary equation of motion. Motivated by the duality between the wave-scattering problem and the point-mass scenario discussed in Sec.~\ref{sec:scat}, we compute the energy of a point mass orbiting a compact star, following a self-force approach similar to that of Ref.~\cite{Feng:2024olt}. Below we briefly summarize the basic strategy; the detailed formalism is presented in Appendix~\ref{sec:pointmass}.

The metric of a point particle $m_2$ orbiting the compact star $m_1$ with $m_2\ll m_1$ can be written as
\begin{equation}
    g_{ab} = g^0_{ab} +h_{ab}\,,
\end{equation}
where $g^0_{ab}$ is the metric of the static background spacetime and $h_{ab}$ is the metric perturbation.  
Without loss of generality, we consider that the circular orbit of the point mass is confined to the equatorial plane, $x^a(\tau) = [t(\tau), r_0, \theta_0=\frac{\pi}{2}, \phi(\tau)]$. We then solve the metric perturbation in Regge-Wheeler gauge. 
With the perturbations at hand, we can locally define a mechanical binding energy based on the first law of binary mechanics (FLBM)~\cite{LeTiec:2011ab,LeTiec:2011dp}, which expresses the binding energy in terms of the first-order Detweiler redshift $z_{\rm SF}$~\cite{Detweiler:2008ft},
\begin{equation}
    E_{\rm SF} = \frac{1}{2}z_{\rm SF}-\frac{y}{3}\frac{dz_{\rm SF}}{dy}-1+\sqrt{1-3y}+\frac{y}{6}\frac{5-12y}{(1-3y)^{3/2}}\,,
\end{equation}
where
\begin{equation}
    z_{\rm SF} = -(1-3y)^{1/2}\frac{1}{2}\bar u_a\bar u_b h^{\rm R}_{ab}\,,
\end{equation}
is gauge-invariant and $y\equiv(m_1\Omega)^{2/3}$. 
Here $\bar u^a$ is the non-radial part of the four-velocity of the particle with no $h_{ab}$ terms. 
$h^{\rm R}_{ab}$ denotes the regular part of the metric perturbation. 
Note that in our case, the Detweiler redshift can be decomposed into two parts:
\begin{equation}
    z_{\rm SF} = z_{\rm SF,BH} +z_{\rm SF,tid}\,,
\end{equation}
where the first term, $z_{\rm SF,BH}$, corresponds to self-force contribution in binary black hole systems, while the second term, $z_{\rm SF,tid}$, encapsulates the tidal effects arising from the compact star. The self-force contribution $z_{\rm SF, BH}$ has been extensively studied in the literature~\cite{Detweiler:2008ft,LeTiec:2011dp}. In this work, we focus on the tidal contribution, $z_{\rm SF,tid}$, which can be expressed as
\begin{equation}\label{eq:zt}
    z_{\rm SF,tid} = -(1-3y)^{1/2}\frac{1}{2}\bar u_a\bar u_b \left(h_{ab}^{\rm CS}-h_{ab}^{\rm BH}\right)\,.
\end{equation}
Since the singular parts of $h^{\rm CS}_{ab}$ and $h^{\rm BH}_{ab}$ are identical in our setup, no regularization of $h_{ab}$ is needed. 

It is important to note that in SF calculations, the redshift is viewed as an expansion in the mass ratio $q\equiv m_2/m_1$, evaluated at fixed $y$. For comparison with PN and EOB results, it is convenient to switch to the standard dimensionless invariant PN parameter, $x\equiv[(m_1+m_2)\Omega]^{2/3}$. In terms of $x$, the Detweiler redshift at linear order in the symmetric mass ratio $\nu\equiv m_1m_2/(m_1+m_2)^2$ becomes
\begin{equation}
    z(x) =\frac{x}{\sqrt{1-3x}}+z_{\rm SF,BH}(x)+z_{\rm SF, tid}(x)\,.
\end{equation}

We derive in Appendix~\ref{sec:analytic} the analytic solution of the $(\ell,m)=(2,0)$ mode contribution to the Detweiler redshift [Eq.~\eqref{eq:z20 7PN}], denoted as $z_{\rm SF,tid}^{20}$. The contributions from the $(\ell, m)=(2,\pm2)$ modes, $z_{\rm SF,tid}^{22}$, can be obtained numerically. 
In what follows, we provide an analytic representation of the numerical data in the following form:
\begin{equation}\label{eq:z22 fit}
\begin{split}
    z_{\rm SF,tid}^{22}(x) =& z_{\rm SF,tid}^{20}\frac{\omega_f^2}{\omega_f^2-4\Omega^2}(\alpha_0+\alpha_1x+\alpha_2x^2\\
    &+\alpha_3x^3+\alpha_4x^4+\alpha_5x^5+\alpha_6x^6)\,,
\end{split} 
\end{equation}
with $\alpha_0=3/2$, which captures both the asymptotic behavior $z_{\rm SF,tid}^{22} \to \frac{3}{2} z_{\rm SF,tid}^{20}$ at $x \to 0$, and the resonance feature at $2\Omega \to \omega_f$. 
The coefficients are found to be $\alpha_1 = -6.005$ and $\alpha_2 = 21.94$, with uncertainty in the last digit. Therefore, in the following, we fix $\alpha_1 = -6$ and $\alpha_2 = 22$, while all other coefficients $\alpha_i$ ($i\geq3$) are obtained by fitting to numerical data. The optimal coefficients are found to be 
\begin{subequations}
\begin{align}
  \alpha_3 &= -60.18736362\,, \\
  \alpha_4 &=-635.63344683\,, \\
  \alpha_5 &= 5931.89992854\,,\\
  \alpha_6 &= -18416.0079901\,.
\end{align}
\end{subequations}

\subsection{Effective-one-body approach}\label{sec:EOB}
In this subsection, we implement the tidal contribution of the self-force into the EOB formalism, following the approach first introduced in Ref.~\cite{Barausse:2011dq}.
Within the EOB framework, the two-body dynamics of non-spinning compact objects is described by mapping the real Hamiltonian $H_{\rm EOB}$ to an effective Hamiltonian $H_{\rm eff}$, which corresponds to a test particle of mass $\mu = \frac{m_1 m_2}{m_1 + m_2}$ moving in a deformed Schwarzschild background with total mass $M = m_1 + m_2$, via~\cite{Buonanno:1998gg}
\begin{equation}
    H_{\rm EOB} = M\sqrt{1+2\nu\left(\frac{H_{\rm eff}}{\mu}-1\right)}\,.
\end{equation}
The EOB effective metric reads
\begin{equation}
    g_{ab}^{\rm eff}dx^ax^b= -A(r)dt^2+B(r)dr^2+r^2d\Omega^2\,,
\end{equation}
together with the mass-shell constraint~\cite{Damour:2000we}
\begin{equation}
    g_{\rm eff}^{ab}p_{a}p_{b}+\mu^2+Q=0\,.
\end{equation}
Here, the potential $Q$ encodes possible effective interactions that lead to a non-geodesic motion. 
The effective Hamiltonian is given by
\begin{equation}
    H_{\rm eff}^2 = A\left(\mu^2+B^{-1}p_r^2+p_\phi^2u^2+Q\right)\,.
\end{equation}
It is convenient to work in the DJS gauge~\cite{Damour:2000we}, in which $Q$ depends only on the radial momentum $p_r$, such that $Q = 0$ for the circular orbits considered in this work. 
The EOB potential $A(r)$ can be written as~\cite{Buonanno:1998gg,Damour:2000we,Barausse:2011dq}
\begin{align}
    A(u) &= A_{\rm pm}(u)+ A_{\rm tid}(u)\,,\\
    A_{\rm pm}(u) &= 1-2u+2\nu u^3+\left(\frac{94}{3}-\frac{41}{32}\pi^2\right)\nu u^4\,,
\end{align}
where $u\equiv M/r$ denotes the inverse Schwarzschild-like EOB radial coordinate.
With the potential $A(u)$ at hand, we can compute the circular-orbit EOB energy as
\begin{align}
    E_{\rm EOB}(u) &= M\sqrt{1+2\nu\left(\frac{E_{\rm eff}}{\mu}-1\right)}\,,\\
    E_{\rm eff}(u) &= \mu\sqrt{\frac{2A^2(u)}{2A(u)+uA^\prime(u)}}\,.
\end{align}
By computing the difference between EOB energy with and without the tidal contribution $A_{\rm tid}$, we can obtain the tidal energy as function of orbit frequency $\Omega$, 
\begin{equation}
    E_{\rm tid}(\Omega) = E_{\rm EOB}(\Omega;A=A_{\rm pm}+A_{\rm tid})-E_{\rm EOB}(\Omega;A=A_{\rm pm})\,,
\end{equation}
where 
\begin{equation}
    \Omega = \frac{\partial H_{\rm EOB}}{\partial p_\phi}\,.
\end{equation}

\begin{figure}
    %\centering
    \includegraphics[width=0.75\linewidth]{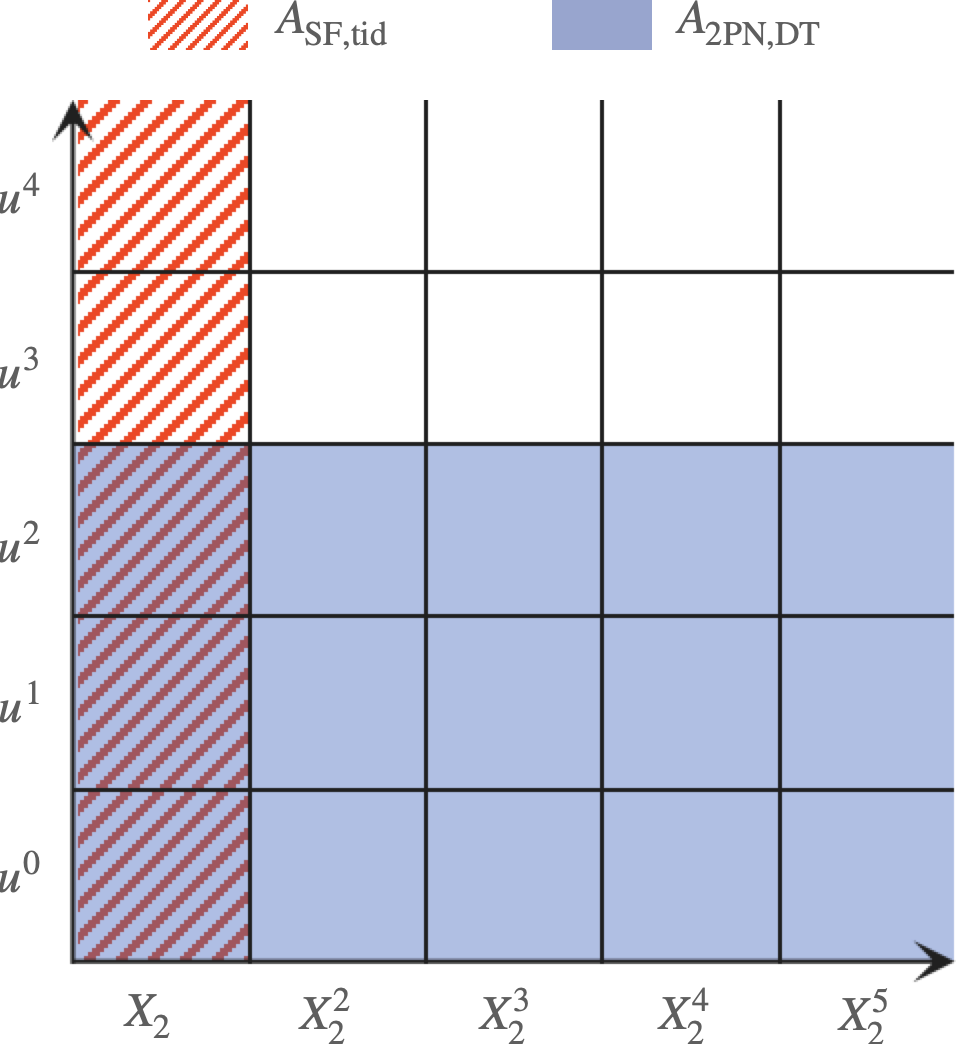}
    \caption{Schematic illustration of $A_{\rm{SF,tid}}$ and $A_{\rm{2PN,DT}}$ as functions expanded in terms of $X_2$ and $u$. The overlapping region, $A_{\rm{ovp}}$, corresponds to the intersection between the red slanted lines and the blue shaded area.}
    \label{fig:hybrid}
\end{figure}

In our self-force formalism, the tidal contribution to the EOB potential, $A_{\rm tid}(u)$, is found to be related to the redshift function via~\cite{Barausse:2011dq}
\begin{equation}\label{eq:Asf}
    A_{\rm SF,tid}(u) =\nu z_{\rm SF,tid}(u)\sqrt{1-3u}+\mathcal{O}(\nu^2)\,.
\end{equation}
Although our current analysis is restricted to linear order in $\nu$, we are able to determine all PN corrections at this order (see, e.g. Fig.~\ref{fig:hybrid}).
On the other hand, the tidal contribution can also be derived from an effective-action approach. In the adiabatic-tide limit (AT), the tidal part of $A(u)$ can be expressed up to 2PN order as~\cite{Bini:2012gu}
\begin{equation}\label{eq:A2PN AT}
\begin{split}
    A_{\rm 2PN,AT} =& -\frac{3\lambda X_2 u^6}{X_1M^5}\bigg[1+\frac{5}{2}X_1 u\\
    &+\left(\frac{337}{28}X_1^2+\frac{1}{8}X_1+3\right)u^2+\mathcal{O}(u^3)\bigg]\,,
\end{split}
\end{equation}
where $X_1=m_1/M$ and $X_2=m_2/M$. $\lambda$ is the tidal deformability. 
It is worth noting that our self-force expression, Eq.~\eqref{eq:Asf}, agrees with Eq.~\eqref{eq:A2PN AT} at 2PN order. To see this, we plug Eq.~\eqref{eq:z20 7PN} and Eq.~\eqref{eq:z22 fit} into $z_{\rm SF,tid}=z_{\rm SF,tid}^{20}+2z_{\rm SF,tid}^{22}$ and take the adiabatic limit. This yields
\begin{equation}
    A_{\rm SF,AT} = -\frac{3\lambda X_2u^6}{M^5}\left[1+\frac{5}{2}u+\frac{849}{56}u^2+\mathcal{O}(u^3)\right]\,,
\end{equation}
which is in exact agreement with Eq.~\eqref{eq:A2PN AT} when taking $X_1\to 1$.
To capture the dynamic tidal effects, Ref.~\cite{Steinhoff:2016rfi} suggests an effective approach by replacing the static tidal deformability $\lambda$ in Eq.~\eqref{eq:A2PN AT} with $\lambda_{\rm eff}$, i.e.,
\begin{equation}\label{eq:A2PN DT}
\begin{split}
    A_{\rm 2PN,DT} =& -\frac{3\lambda_{\rm eff} X_2 u^6}{X_1M^5}\bigg[1+\frac{5}{2}X_1 u\\
    &+\left(\frac{337}{28}X_1^2+\frac{1}{8}X_1+3\right)u^2+\mathcal{O}(u^3)\bigg]\,,
\end{split}
\end{equation}
where 
\begin{equation}
    \lambda_{\rm eff} = \frac{\lambda}{4}+\frac{3\lambda}{4}\frac{\omega_f^2}{\omega_f^2-4\Omega^2}\,.
\end{equation}

We now have two independent expressions for the tidal part of the EOB potential, $A_{\rm tid}$, given by Eq.~\eqref{eq:Asf} and Eq.~\eqref{eq:A2PN DT}, both of which can be expanded as series in $X_2$ and $u$, as illustrated in Fig.~\ref{fig:hybrid}. Inspired by Ref.~\cite{Feng:2021sax}, we propose to construct a hybrid tidal potential, which combines information from both approaches,
\begin{equation}
    A_{\rm tid} = A_{\rm SF,tid}+A_{\rm 2PN,DT}-A_{\rm ovp}\,,
\end{equation}
where
\begin{equation}
    A_{\rm ovp} = -\frac{3\lambda_{\rm eff}X_2 u^6}{M^5}\left(1+\frac{5}{2}u+\frac{849}{56}u^2\right)
\end{equation}
denotes the contribution from the overlap regime as shown in Fig.~\ref{fig:hybrid}.

\begin{figure}
    %\centering
    \includegraphics[width=0.9\linewidth]{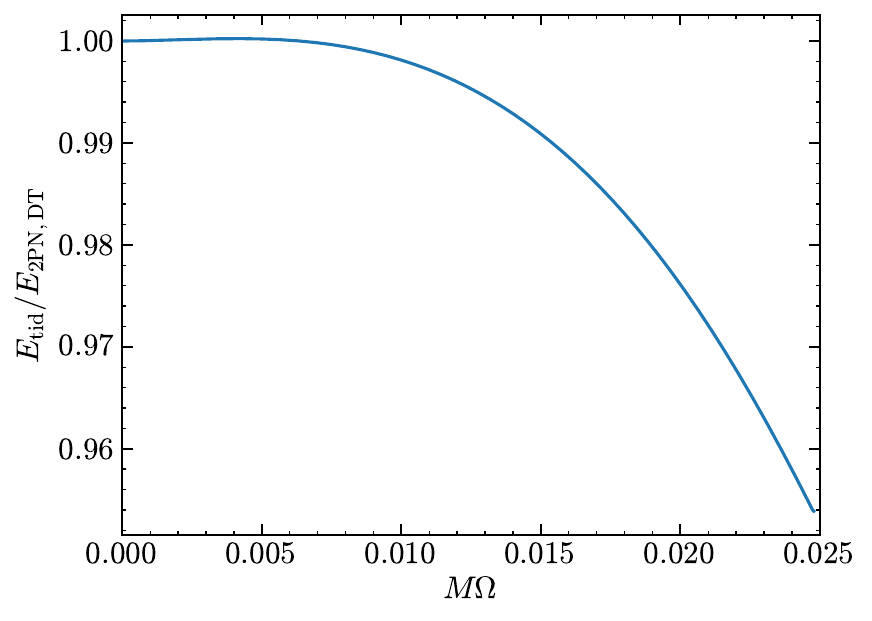}
    \caption{Ratio of tidal energy computed using $A_{\rm tid}$ and $A_{\rm 2PN,DT}$ in the extreme mass-ratio limit ($m_2 \ll m_1$). The stellar model is constructed from a polytropic equation of state with $n = 1$, and fixed to $M=m_1=1.4\,M_\odot$ and $\mathcal{C}=0.15$.}
    \label{fig:etid_extremeq}
\end{figure}
\begin{figure}
    %\centering
    \includegraphics[width=0.9\linewidth]{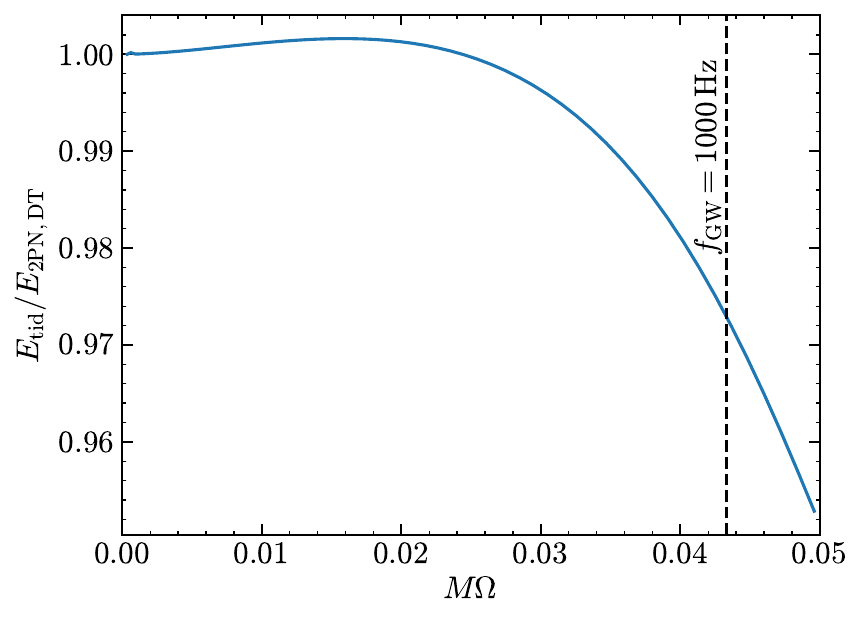}
    \caption{Same as Fig.~\ref{fig:etid_extremeq}, but for a equal-mass binary ($M=m_1+m_2=2.8\,M_\odot$). The dashed line indicates the GW frequency of $f_{\rm GW} =\Omega/\pi=1000\,{\rm Hz}$. }
    \label{fig:etid_equalmass}
\end{figure}

\begin{figure}
    %\centering
    \includegraphics[width=0.9\linewidth]{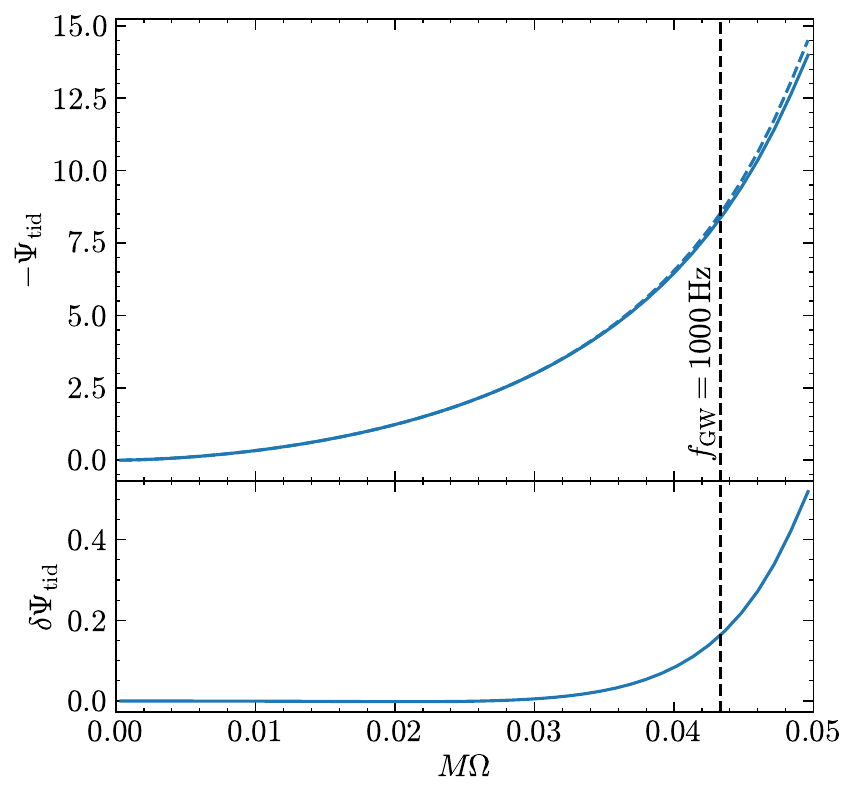}
    \caption{Upper: Accumulated tidal phase computed using $A_{\rm tid}$ (solid line) and $A_{\rm 2PN,DT}$ (dashed) for the equal-mass binary case as shown in Fig.~\ref{fig:etid_equalmass}. Lower: The phase difference between the two models.}
    \label{fig:phase}
\end{figure}

\subsection{Results}\label{sec:res}
In Fig.~\ref{fig:etid_extremeq} and Fig.~\ref{fig:etid_equalmass}, we compare the tidal energy computed from $A_{\rm tid}$ and $A_{\rm 2PN,DT}$ in the extreme mass-ratio limit and in the equal-mass case, respectively. 
The compact star is constructed from a polytropic equation of state with $n = 1$,
and fixed to $m_1= 1.4\,M_\odot$ and $\mathcal{C}=0.15$ ($R=13.83\,{\rm km}$), corresponding to a dimensionless tidal deformability $\Lambda_1\equiv \lambda/m_1^5=715$. 
In both cases, we find that incorporating self-force corrections leads to a $\sim 3\%$ correction in the tidal energy at a GW frequency of $f_{\rm GW}=\Omega/\pi=1000\,{\rm Hz}$.

To calculate the GW phase difference accumulated during the inspiral stage, we adopt the following approximation
\begin{equation}\label{eq:phase}
    \frac{d\Psi_{\rm tid}}{d\Omega } =  2\Omega\frac{dE_{\rm tid}/d\Omega}{\dot E}\,,
\end{equation}
where 
\begin{equation}
    \dot E = -\frac{32}{5}M^{4/3}\mu^2\Omega^{10/3}\,.
\end{equation}

In Fig.~\ref{fig:phase}, we show the accumulated tidal phase obtained using $A_{\rm tid}$ model and $A_{\rm 2PN,DT}$ model. It can be seen that incorporating self-force corrections leads to a phase modulation reaching approximately $0.2$ radian at $f_{\rm GW} \sim 10^3\,{\rm Hz}$, and over $0.5$ radian before the merger.

\section{Conclusion}\label{sec:conclusion}

In this work, we study the excitation of compact stars under external tidal fields in the fully relativistic regime. This problem has been studied in the Newtonian regime in the past, where the external tidal forces can be decomposed into a mode basis and the star oscillations can be effectively described by the evolution of a set of modes. For relativistic stars, this framework no longer holds because the quasi-normal modes of the star do not form a complete basis for star perturbations, nor is an orthogonal projection of modes known.

We find that the scattering wave formalism becomes particularly useful in studying this relativistic problem. First, it can be shown that a system with a point mass orbiting around the star can be mapped to another system with purely input and output waves. The scattering formulation naturally provides a gauge-invariant way to measure the strength of the external tidal field, which can be quantified by the amplitude of the ingoing wave in the equivalent scattering problem. Second, we have shown that the response function defined in the scattering formulation encodes critical information about the tidal response of the star. For example, the poles of the response function correspond to quasi-normal modes of the star, and the residue of poles has a direct analogy with the {\it black hole excitation factor} in black hole perturbation theory~\cite{Leaver:1986gd}, which measures the susceptibility of modes to external drivings. It turns out that the residue ${\rm W}_n$ is approximately by the decay rate of the particular mode. In the equilibrium tide scenario, the phase of the response is intimately related to the total tidal energy within the spacetime, including both the fluid and gravitational perturbations.

With a time-dependent external tidal field, a set of modes may be transiently excited if the tidal field frequency crosses the mode frequency at certain times. We are able to compute the energy of the transiently excited modes during this process, which is proportional to $|{\rm W}_n|$ and the square of the amplitude of the external tidal field. 
The relativistic formula agrees with the Newtonian formula in the low-frequency, weak-gravity regime, but differs significantly for compact stars. For example, for a $1.4\,M_\odot$ neutron star with radius $\sim 13\,{\rm km}$, if a g$_1$-mode is resonantly excited, the Newtonian formula may over-predict the mode energy by a factor of ten. This large discrepancy calls for necessary re-examination of previous resonant mode calculation for various kinds of modes. Implementing Eq.~\eqref{eq:mode energy} may have a dramatic impact on the predictions.

The $\ell=2$ f-mode represents the dominant contribution of equilibrium tide in inspiraling binary neutron star systems.  Because the f-mode frequency is generally larger than the merger frequency, these modes are not expected to be on resonance in the inspiral stage. Within the scattering formalism, through a flux-balance calculation, we can prove that the tidal energy in spacetime is proportional to $\partial \theta/\partial \omega$, where $\theta$ is the phase of the response function.  This formula is numerically checked with the Hamiltonian formula of perturbed stars derived by Monrief in~\cite{Moncrief:1974am,Moncrief:1974an}. However, this tidal energy cannot be directly used to compute the tidal binding energy of a binary neutron star system, as part of the energy computed in the scattering formalism belongs to the freely propagating waves.

In order to compute the tidal energy of a binary star system in the relativistic regime, we study the model problem with a point mass orbiting around a star. The tidal response of the star formally appears at the same order as the gravitational self force of the point mass in the expansion of the mass ratio. In particular, in order to account for the tidal energy of the binary, it is best to compute the gravitational field at infinity to extract the ADM mass of the whole system, which formally requires gravitational perturbations to the sub-leading order in mass-ratio (the same order as the second-order self force). Fortunately according to the discussion in~\cite{LeTiec:2011ab}, one can apply the first law of binary mechanics to obtain a canonical energy that only depends on the first-order gravitational perturbation in the near-zone. For the black hole case the definition with canonical energy agrees with the definition using ADM mass within $10^{-5}$ fractional difference~\cite{Pound:2019lzj}. Therefore we also apply this definition of canonical energy to approximate the tidal energy of a star binary. In the extreme mass ratio limit, this energy differs the effective-one-mode energy that has been widely adopted in literature by up to $3\%$ at $f_{\rm GW}\sim10^3\,{\rm Hz}$. Adopting an Effective-One-Body formalism to scale this canonical energy for comparable mass-ratio systems, we find that the resulting phase modulation for binary neutron stars can reach $0.2\,{\rm rad}$ at $10^3\,{\rm Hz}$ and $\ge 0.5\,{\rm rad}$ before merger. Note that the phase modulation only includes the relativistic correction on the tidal energy, but there is also possibly a correction on the tidal flux. Together the phase modulation may be significant, so that this new relativistic treatment of tidal energy should be incorporated into future waveforms of neutron star binaries.

As the binary enters the late inspiral stage, the excitation of stars may receive additional contributions beyond the equilibrium-tide prediction, e.g. the terms following the first term in Eq.~(6.50) of Ref.~\cite{Steinhoff:2016rfi}. These terms directly correspond to ``free oscillations'' of the star upon full excitation of the mode across the resonance. According to the analysis in Sec.~\ref{sec:transient}, the relativistic energy correction of transiently excited f-modes may be greater than $50\%$. Even if the f-mode is not fully excited in the inspiral stage, the relativistic correction for this part of the contribution can still be significant.  It is also important to note that the calculations in Sec.~\ref{sec:eqt} assume an equilibrium tide scenario, so that they are not directly applicable to describe non-steady effects. However, it may be instructive to explore the scenario with a point mass orbiting around a star following an inspiralling orbit, which may be mapped to these systems with non-equilibrium tides. This study is beyond the scope of this work.

%%%%%%%%%%%%%%%%%%%%%%%%%%%%%%%%%%%%%%%%%%%%%%%%%%%%%%%%%%%%%%
%%%%%%%%%%%%%%%%%%%%%%%%%%%%%%%%%%%%%%%%%%%%%%%%%%%%%%%%%%%%%%
\section{Acknowledgements} 
\label{sec:acknowledgements}
\appendix

We are grateful to Zihan Zhou for answering questions about scattering amplitude calculations of neutron star tides and Sizheng Ma for discussions about the scattering formalism. We thank Eric Poisson for many useful comments on the manuscript. ZM is supported by the Postdoctoral Innovation Talent Support Program of CPSF (No. BX20240223) and CPSF funded project (No. 2024M761948).

\section{Perturbation of relativistic stars}\label{sec:appendix pert}
In this Appendix, we describe the technical details involved in the calculation of perturbation of relativistic stars. 

\subsection{Static background spacetime}\label{subsec:appendix bg}
For a non-rotating star, the static and spherically symmetric line element is given by
\begin{equation}
	ds^2 =g_{ab}dx^a dx^b= -e^{2\nu}dt^2+e^{2\lambda}dr^2+r^2(d\theta^2+\sin^2\theta d\phi^2)\,,
\end{equation}
where the metric components $\nu$ and $\lambda$ are functions of $r$ only. We model the star as a perfect fluid, with the stress-energy-momentum tensor given by
\begin{equation}
    T^{ab} = (p+\varepsilon)u^a u^b+pg^{ab}\,,
\end{equation}
where $u^a$ is the four-velocity. $p$ and $\varepsilon$ denote the pressure and energy density, respectively, which are connected via an equation of state $p = p(\varepsilon)$.
The Einstein field equations for this system reduce to the Tolman–Oppenheimer–Volkoff (TOV) equations, expressed as
\begin{align}
    \frac{dm}{dr} &= 4\pi r^2\varepsilon\,,\label{eq:dmdr}\\
    \frac{dp}{dr} &= -\frac{(p+\varepsilon)(m+4\pi r^2p)}{r(r-2m)}\,,\label{eq:dpdr}\\
    \frac{d\nu}{dr}&=\frac{m+4\pi r^2p}{r(r-2m)}\,.\label{eq:dnudr}
\end{align}
where $m(r)$ is defined as
\begin{equation}
    m(r) = \frac{r}{2}\left[1-e^{-2\lambda(r)}\right]\,.
\end{equation}
Given the central pressure $p_0$ (or energy density $\varepsilon_0$) at $r=0$, Eqs.~(\ref{eq:dmdr}--\ref{eq:dnudr}) can be integrated outward the pressure vanishes, i.e., $p(R)=0$.
This determines the stellar radius $R$, and the corresponding gravitational mass $M = m(R)$. 
Note there is an arbitrary constant in the function $\nu$, which can be fixed by requiring 
\begin{equation}
    \nu(R) = -\lambda(R) = \frac{1}{2}\ln\left(1-\frac{2M}{R}\right)\,.
\end{equation}

\subsection{Perturbations inside the star}\label{subsec:appendix pert inside}
In this work we only consider the even-parity perturbations. We write the metric perturbation in Regge-Wheeler gauge as~\cite{Regge:1957td,Lindblom:1983ps}
\begin{equation}\label{eq:metric pertur}
\begin{split}
    h_{ab} = & \big[\zeta^\ell H_0e^{2\nu}dt^2+2i\omega r\zeta^{\ell}H_1dtdr+\zeta^\ell H_0e^{2\lambda}dr^2 \\
    & +\zeta^\ell Kr^2(d\theta^2+\sin^2\theta d\phi^2)\big]Y_{\ell m}(\theta,\phi)e^{-i\omega t}\,,
\end{split}
\end{equation}
with 
\begin{equation}
\zeta= \left\{\!
\begin{aligned}
&r/R\,, & 0\leq r <R\,, \\ 
&1\,, & r\geq R\,,
\end{aligned}
\right.
\end{equation}
and $Y_{\ell m}(\theta,\phi)$ the spherical harmonics.
We also write the fluid displacement as
\begin{align}
    \xi^r = &\zeta^\ell r^{-1}e^{-\lambda}W Y_{\ell m}e^{-i\omega t}\,,\\
	\xi^\theta = &-\zeta^\ell r^{-2}V \partial_\theta Y_{\ell m}e^{-i\omega t}\,,\\
	\xi^\phi = &-\zeta^\ell r^{-2}\sin^{-2}\theta V\partial_\phi Y_{\ell m}e^{-i\omega t}\,.
\end{align}
The Eulerian perturbations of energy density and pressure are given by
\begin{align}
    \delta \varepsilon &= (p+\varepsilon)\frac{\Delta n}{n}-\xi^i\frac{d\varepsilon}{dx^i}\,,\label{eq:p pertur}\\
    \delta p &= \Gamma p\frac{\Delta n}{n}-\xi^i\frac{dp}{dx^i}\,,\label{eq:e pertur}
\end{align}
where $\Gamma = [(p+\varepsilon)/p](\partial p/\partial \varepsilon)_{s}$ is the adiabatic index. The Lagrangian perturbation of the baryon number density $\Delta n$ can be derived from the baryon number conservation $\nabla_\mu(nu^\mu)=0$, it reads
\begin{equation}
    \Delta n = -n\nabla_i\xi^i-\frac12n\delta g^{(3)}/g^{(3)},
\end{equation}
with $g^{(3)}$ the determinant of the metric of the 3-geometry at constant time.
Plugging these perturbations [Eqs.~(\ref{eq:metric pertur}-\ref{eq:e pertur})] into linear Einstein equation $\delta G_{\mu\nu}=8\pi\delta T_{\mu\nu}$, we get~\cite{Lindblom:1983ps}
\begin{widetext}
\begin{align}
    H_1^\prime &=\left(\lambda^\prime-\nu^\prime-\frac{\ell+1}{r}\right)H_1-\frac{e^{2\lambda}}{r}\left[H_0+K+16\pi(p+\varepsilon)V\right]\,,\label{eq:H1}\\
    K^\prime &= \left(\nu^\prime-\frac{\ell+1}{r}\right)K+\frac{1}{r}H_0-\frac{\ell(\ell+1)}{2r}H_1+8\pi(p+\varepsilon)\frac{e^{\lambda}}{r}W\,,\label{eq:K}\\
    W^\prime &=-\frac{\ell+1}{r}W+ re^{\lambda}\left[\frac{e^{-\nu}}{\Gamma p}X-\frac{\ell(\ell+1)}{r^2}V-\frac12 H_0-K\right]\,,\label{eq:W}\\
    X^\prime &=-\frac{\ell}{r}X+ (p+\varepsilon)e^{\nu}\Bigg\{\left(\frac{\nu^\prime}{2}-\frac{1}{2r}\right)H_0+\left(\frac{1}{2r}-\frac{3\nu^\prime}{2}\right)K+\left[\frac{1}{2}\omega^2re^{-2\nu}+\frac{\ell(\ell+1)}{4r}\right]H_1
    -\frac{\ell(\ell+1)}{r^2}\nu^\prime V \nonumber\\
    &-\frac{1}{r}\left[\omega^2e^{\lambda-2\nu}+4\pi(p+\varepsilon)e^{\lambda}-r^2\left(\frac{e^{-\lambda}}{r^2}\nu^\prime\right)^\prime\right]W\Bigg\}\,,\label{eq:X}
\end{align}
\end{widetext}
with 
\begin{widetext}
\begin{align}
    H_0=&\Bigg\{-8\pi r^2e^{-\nu}X-\left[\omega^2r^2e^{-2\nu}-\frac{1}{2}(\ell-1)(\ell+2)+r^2e^{-2\lambda}\left(\nu^{\prime2}-\frac{\nu^\prime}{r}\right)\right]K\nonumber \\
    &-e^{-2\lambda}\left[\omega^2r^2e^{-2\nu}-\frac{1}{2}\ell(\ell+1)r\nu^\prime \right]H_1 \Bigg\}\left[(r\nu^\prime-1) e^{-2\lambda}+\frac{1}{2}\ell(\ell+1)\right]^{-1}\,, \\
    X =& \omega^2(p+\varepsilon)e^{-\nu}V-\frac{e^{\nu-\lambda}}{r}p^\prime W -\frac12 (p+\varepsilon)e^{\nu}H_0\,,
\end{align}
\end{widetext}
where a prime denotes differentiation with respect to $r$.
At the center of the star ($r=0$), all perturbation functions must be regular, leading to the boundary conditions,
\begin{align}
    X(0) =&(p_0+\varepsilon_0)e^{\nu(0)}\Bigg\{\bigg[\frac{4\pi}{3}(\varepsilon_0+3p_0)\nonumber\\
    &-\frac{\omega^2e^{-2\nu(0)}}{\ell}\bigg]W(0)-\frac{1}{2}K(0)\Bigg\}\,, \label{eq:center BC1}\\
    H_1(0) =& \frac{-2\ell K(0)+16\pi(p_0+\varepsilon_0)W(0)}{\ell(\ell+1)}\,.\label{eq:center BC2}
\end{align}
At the star surface ($r=R)$, the Lagrangian pressure variation must vanish, imposing another boundary condition as
\begin{equation}\label{eq:surface BC}
    X(R) = 0\,.
\end{equation}
At this point, we have four equations [Eqs.~(\ref{eq:H1}--\ref{eq:X})] governing the radial evolution of four independent variables ($H_1$, $K$, $W$ $X$), along with three constraint equations [Eqs.~(\ref{eq:center BC1}--\ref{eq:surface BC})]. This leaves one remaining degree of freedom, which, as we will see, can be determined by the gravitational perturbation in the exterior spacetime of the star.

\subsection{Perturbations outside the star}\label{subsec:appendix pert outside}
Outside the star, only gravitational perturbations are relevant, where the system of perturbation equations [Eqs.~(\ref{eq:H1}--\ref{eq:X})] reduce to a second-order wave function~\cite{Zerilli:1970se}
\begin{equation}\label{eq:Zerilli eq}
    \frac{d^2 Z}{dr_*^{2}}+(\omega^2-V_Z)Z = 0\,,
\end{equation}
with the effective potential given by
\begin{equation}
    V_Z = \frac{e^{-2\lambda}\left[2n^2(n+1)r^3+6n^2Mr^2+18nM^2r+18M^3\right]}{r^3(nr+3M)^2}\,.
\end{equation}
Here $n=(\ell-1)(\ell+2)/2$ and the tortoise coordinate $r_*$ is defined as 
\begin{equation}
    \frac{d}{dr_*} = e^{\nu-\lambda}\frac{d}{dr}\,.
\end{equation}

The master variable $Z$ is related to the metric perturbation functions as
\begin{align}
    Z(r_*) =& \frac{r^2}{nr+3M}e^{-2\lambda}H_1+\frac{r^2}{nr+3M}K\,,\label{eq:Z_KH1}\\
    \frac{dZ(r_*)}{dr_*} =& -\frac{n(n+1)r^2+3nMr+6M^2}{(nr+3M)^2}e^{-2\lambda}H_1\nonumber\\
    &-\frac{nr^2-3nMr-3M^2}{(nr+3M)^2}K\,.\label{eq:dZ_KH1}
\end{align}
On the surface of the star, it must match the internal solutions of $H_1(R)$ and $K(R)$ obtained by solving Eqs.~(\ref{eq:H1}--\ref{eq:X}).

Eq.~\eqref{eq:Zerilli eq} admits two linearly independent solutions with the following asymptotic behavior at large $r$, 
\begin{align}
    Z_{\rm out} =& e^{i\omega r_*} \left[1+a_1r^{-1}+a_2r^{-2}+a_3r^{-3}+\mathcal{O}(r^{-4})\right]\,,\label{eq:Zout series}\\
    Z_{\rm in} =& e^{-i\omega r_*} \left[1+a_1^*r^{-1}+a_2^*r^{-2}+a_3^*r^{-3}+\mathcal{O}(r^{-4})\right]\,,\label{eq:Zin series}
\end{align}
which corresponds to the outgoing and ingoing wave, respectively.
The coefficients $a_j$ and their complex conjugates $a_j^*$ are given by 
\begin{align}
    a_1 =& \frac{i(n+1)}{\omega}\,,\\
    a_2 =& -\frac{1}{2\omega^2}\left[n(n+1)+\frac{3i(n+2)}{n}M\omega\right]\,,\\
    a_3 =& -\frac{i}{6\omega^3}\bigg[n(n-2)(n+1)+3i(n+4)M\omega\nonumber\\
    &-\frac{18(2n+3)}{n^2}M^2\omega^2\bigg]\,.
\end{align}
%Accordingly,
The general solution to Eq.~\eqref{eq:Zerilli eq} can be expressed as 
\begin{equation}\label{eq:Z solution}
    Z = A_{\rm out} Z_{\rm out} +A_{\rm in}Z_{\rm in}\,.
\end{equation}
The coefficients $A_{\rm out}$ and $A_{\rm in}$ represent the amplitudes of the outgoing and ingoing waves, respectively.
To obtain $A_{\rm out}$ and $A_{\rm in}$, we directly integrate Eq.~\eqref{eq:Zerilli eq} and then match the results to Eq.~\eqref{eq:Z solution} with the asymptotic expansion in Eqs.~(\ref{eq:Zout series} \& \ref{eq:Zin series}) at large $r$. In practice, we set $r=300/\omega$ to achieve a relatively high precision solution.

\section{Computing the poles of $\mathcal{R}$ and the corresponding residue ${\rm W}_n$}\label{sec:appendix residue}
Given any frequency $\omega$, the ingoing and outgoing amplitudes $A_{\rm in}$ and $A_{\rm out}$ can be obtained using the method presented in Appendix~\ref{sec:appendix pert}, which enables us to compute the scattering/reflection coefficient
\begin{equation}
    \mathcal{R}(\omega) = \frac{A_{\rm out}}{A_{\rm in}}\,.
\end{equation}

The function $\mathcal{R}(\omega)$ possesses series poles in the complex frequency plane, which correspond to the quasi-normal modes of the star. 
To compute these poles, we proceed as follows:
\begin{enumerate}
    \item We first search along the real axis to obtain an approximate value of the real part of the mode frequency, $\omega_{n{\rm R}}^{1}$, following the method described in Ref.~\cite{Lindblom:1983ps}.
    \item We then search along the imaginary axis to find the frequency $\omega_{n{\rm R}}^{1} + i\gamma_n^{1}$ that maximizes $|\mathcal{R}|$.
    \item Next, we search again along the real axis for the frequency $\omega_{n{\rm R}}^{2} + i\gamma_n^{1}$ that maximizes $|\mathcal{R}|$.
    \item Steps 2 and 3 are repeated iteratively until the desired precision for the frequency is achieved.
\end{enumerate}

In the actual calculation, the internal solution of the perturbation contains numerical errors arising from interpolation errors in the background star model. These errors can affect the precise determination of $\mathcal{R}$. To overcome this issue, we first compute a set of $H_1(R,\omega)$ and $K(R,\omega)$ in a small interval $[\omega_{n{\rm R}}^1-\Delta\omega,\omega_{n{\rm R}}^1+\Delta\omega]$ at star surface $r=R$. We then use $(5,5)$ Padé approximation to fit these functions in this region, i.e, 
\begin{align}
    H_1(R,\tilde\omega) = &\frac{A_0+A_1\tilde\omega+A_2\tilde\omega^2+A_3\tilde\omega^3+A_4\tilde\omega^4+A_5\omega^5}{B_0+B_1\tilde\omega+B_2\tilde\omega^2+B_3\tilde\omega^3+B_4\tilde\omega^4+B_5\tilde\omega^5}\\
    K(R,\tilde\omega) = &\frac{C_0+C_1\tilde\omega+C_2\tilde\omega^2+C_3\tilde\omega^3+C_4\tilde\omega^4+C_5\tilde\omega^5}{D_0+D_1\tilde\omega+D_2\tilde\omega^2+D_3\tilde\omega^3+D_4\tilde\omega^4+D_5\tilde\omega^5}
\end{align}
where $\tilde\omega =(\omega-\omega_{nR}^1)/\Delta\omega$.
This allows us to ensure the smoothness of the internal solution at the star’s surface, thereby reducing errors when solving for the external solution. With this method, we can achieve high precision in calculating the f-mode and g-mode frequencies (including their imaginary part).

Once the pole $\omega_n = \omega_{nR}+i\gamma_n$ is located, the residue ${\rm W}_n$ can be directly calculated as follows
\begin{equation}
    {\rm W}_n = \left[{\frac{d}{d\omega}\left(\frac{1}{\mathcal{R}}\right)}\right]^{-1}\bigg|_{\omega=\omega_n}\,.
\end{equation}
However, this method can be numerically challenging, as $\mathcal{R}$ diverges near the pole, which can lead to significant loss of accuracy in the numerical differentiation.
To overcome this difficulty, an alternative and numerically stable approach is to evaluate the residue via a contour integral around the pole. According to the residue theorem, the residue ${\rm W}_n$ at $\omega_n$ can be expressed as
\begin{equation}
    {\rm W}_n = \frac{1}{2\pi i} \oint_{\mathcal{C}} \mathcal{R}(\omega) \, d\omega\,,
\end{equation}
where $\mathcal{C}$ is a small closed contour enclosing the pole $\omega_n$. This method avoids direct differentiation near a divergent point and is therefore less sensitive to numerical noise. In practice, $\mathcal{C}$ can be taken as a small circle of radius $\epsilon$ around $\omega_n$, and the integral can be approximated using the trapezoidal rule over the complex contour:
\begin{equation}
    {\rm W}_n \approx \frac{1}{2\pi i} \sum_{k=1}^{N} \mathcal{R}(\omega_k) \Delta\omega_k\,,
\end{equation}
where $\omega_k=\omega_n+\epsilon e^{2\pi ik/N}$, and $\Delta\omega_k=\omega_{k+1}-\omega_k$. The value of $\epsilon$ should be small enough to stay within the linear regime, yet large enough to avoid floating-point round-off errors. 
In this work, we set $N=20$ and $\epsilon=\gamma_n/5$ for our calculations. To test the numerical accuracy, we vary $N$ to 30 or $\epsilon$ to $\gamma_n/10$, which results in a change of ${\rm W}_n$ by $\delta|{\rm W}_n|/|{\rm W}_n|\sim\mathcal{O}(10^{-5})$ for f-mode and $\sim\mathcal{O}(10^{-3})$ for g-mode. Therefore, we are confident that our results are accurate to this level.

\section{Point mass orbiting around a compact star}\label{sec:pointmass}
In the point mass scenario, the perturbations are described in a manner similar to the formalism provided in Appendix~\ref{sec:appendix pert}, with the only difference being that we have included the source term representing the point particle in the Zerilli equation, i.e., Eq.~\eqref{eq:Zerilli eq} should be altered as
\begin{equation}\label{eq:Zerilli wt sc}
    \frac{d^2 Z}{dr_*^{2}}+(\omega^2-V_Z)Z = S\,.
\end{equation}
In the case of a circular orbit in the equatorial plane, the source term is given by~\cite{Zerilli:1970wzz}
\begin{equation}
    S= \frac{8\pi}{n+1}\mu\delta(\omega-m\Omega)e^{-2\lambda}\left(\frac{dA}{dr}+B\right)Y^*_{\ell m}\left(\frac{\pi}{2},\phi\right)\,,
\end{equation}
where
\begin{align}
    A =& \frac{re^{-4\lambda}}{nr+3M}\frac{1}{\sqrt{1-3M/r}}\delta(r-r_0)\,,\\
    B=&-\frac{n(n+1)r^2+3nMr+6M^2}{(nr+3M)^2r}\frac{e^{-2\lambda}}{\sqrt{1-3M/r}}\delta(r-r_0)\nonumber\\
    &-\frac{1}{n}\left[\frac{\ell(\ell+1)}{2}-m^2\right]\frac{r\Omega^2}{\sqrt{1-3M/r}}\delta(r-r_0)\,.
\end{align}

The general solution of Eq.~\eqref{eq:Zerilli wt sc} can be expressed as
\begin{equation}
    Z = \alpha Z_{\rm out}+\beta Z_{\rm in} +\int_{R_*}^\infty G(r_*,s_*)S(s_*)ds_*\,.
\end{equation}
Here $Z_{\rm out}$ and $Z_{\rm in}$ are the two homogeneous solutions as in Eqs.~(\ref{eq:Zout series} \& \ref{eq:Zin series}). $\alpha$ and $\beta$ denote the amplitudes of the outgoing and ingoing waves, respectively, corresponding to the $A_{\rm {out}}$ and $A_{\rm {in}}$ used in the main text.
$G(r_*,s_*)$ is the Green's function 
\begin{equation}
\begin{split}
    G(r_*,s_*)=& \frac{1}{W}\big[-Z_{\rm out}(r_*) Z_{\rm in}(s_*)\\
    &+Z_{\rm in}(r_*)Z_{\rm out}(s_*)\big]\Theta(r_*-s_*)\,,
\end{split}
\end{equation}
where $W$ is the Wronskian and $\Theta(x)$ is the Heaviside function. In the point mass scenario, there should be only outgoing wave at spatial infinity, which leading to the boundary condition
\begin{equation}
    Z(r_*)\to \mathcal{A}e^{i\omega r_*}\,,\quad r\to \infty\,.
\end{equation}

\section{Analytic solution for $m=0$ mode}\label{sec:analytic}
For the $m = 0$ mode, we have $\omega = m\Omega = 0$, so the previous method of decomposing the solution into ingoing and outgoing parts no longer applies. Fortunately, we can calculate the analytic solution of perturbed metric functions in this situation.
Here we focus only on the case of $\ell = 2$, but our approach can be readily generalized to arbitrary $\ell\geq3$.
We begin with the $\omega = 0$ equation for $H_0$, which reads~\cite{Hinderer:2007mb}:
\begin{equation}\label{eq:Heq}
\begin{split}
    &H_0^{\prime\prime}+\left[\frac{2}{r}+e^{2\lambda}\left(\frac{2m}{r^2}+4\pi r(p-\varepsilon)\right)\right]H_0^{\prime}\\
    &+\left[-\frac{6e^{2\lambda}}{r^2}+4\pi e^{2\lambda}\left(5\varepsilon+9p+\frac{\varepsilon+p}{dp/d\varepsilon}-4\nu^{\prime 2}\right)\right]H_0=S_H\,.
\end{split}
\end{equation}
where $S_H$ denotes a source term at $r=r_0$. 
Outside the star ($r>R$), Eq.~\eqref{eq:Heq} reduces to 
\begin{equation}\label{eq:Heq external}
    H_0^{\prime\prime}+\left(\frac{2}{r}-2\lambda^\prime\right)H_0^\prime-\left(\frac{6e^{2\lambda}}{r^2}+4\lambda^{\prime 2}\right)H_0=S_H\,.
\end{equation}
We can write the exterior solution as
\begin{equation}
H_0= \left\{\!
\begin{aligned}
&c_1 Q_2^{\,2}\left(\frac{r}{M}-1\right)+c_2 P_2^{\,2}\left(\frac{r}{M}-1\right)\,, & R\leq r \leq r_0\,, \\ 
&d_1Q_2^{\,2}\left(\frac{r}{M}-1\right)\,, & r> r_0\,,
\end{aligned}
\right.
\end{equation}
where $P_2^{\,2}$ and $Q_2^{\,2}$ are the associated Legendre functions. $c_1$, $c_2$ and $d_1$ are coefficients to be determined.
With $H_0$ known, we can obtain both $K$ and $H_1$, via
\begin{equation}\label{eq:Hsol}
\begin{pmatrix}
\frac{3M-r}{r(r-2M)} & -\frac{3}{r} \\
\frac{2r(r-2M)+M(r-3M)}{r(r-2M)} & \frac{3M}{r}
\end{pmatrix}
\begin{pmatrix}
K \\
H_1
\end{pmatrix}  
=
\begin{pmatrix}
K^\prime-\frac{1}{r}H_0\\
\frac{2r+3M}{r}H_0
\end{pmatrix}
\end{equation}
where $K^\prime =H_0^\prime+2\nu^\prime H_0$.
Using Eq.~\eqref{eq:Z_KH1} and Eq.~\eqref{eq:dZ_KH1}
we get the solution of $Z$ and $dZ/dr^*$.

For a compact star, we numerically integrate Eq.~\eqref{eq:Heq} and match the interior solution of $H_0$ and $H_0^\prime$ to the exterior solution given in Eq.~\eqref{eq:Hsol}, while for a black hole, we take $c_1=0$ and $c_2=1$. 
Now we are able to determine the ratio $c_1 / c_2$, which is related to the tidal deformability,
\begin{equation}
    \frac{c_1}{c_2} = \frac{45}{8}\frac{\lambda}{M^5}\,.
\end{equation}
The junction condition at $r=r_0$ can be read off from Eq.~\eqref{eq:Zerilli wt sc},
\begin{align}
    [Z] &= \frac{8\pi\mu}{3\sqrt{1-3M/r_0}}\frac{r_0-2M}{2r_0+3M}Y_{20}^*\left(\frac{\pi}{2},\phi\right)\,,\\
    \left[\frac{dZ}{dr^*}\right] &=\frac{8\pi\mu}{\sqrt{1-3M/r_0}}\frac{M^3+8M^2r_0+4r_0^3}{2r_0^2(2r_0+3M)^2}Y_{20}^*\left(\frac{\pi}{2},\phi\right)\,.
\end{align}
Finally, we get the general solution for metric perturbations. For $R\leq r\leq r_0$, we have 
\begin{widetext}
\begin{align}
    &H_0(r) = -\frac{\pi\mu}{6M^3r_0\sqrt{1-3M/r_0}}Y_{20}^*\left(\frac{\pi}{2},\phi\right)\bigg[14M^3-12M^2r_0+6Mr_0^2\nonumber\\
    &+\left(6M^3-12M^2r_0+9Mr_0^2-3r_0^3\right)\log\left(\frac{r_0}{r_0-2M}\right)\bigg]\,\left[\frac{c_1}{c_2}Q_2^{\,2}\left(\frac{r}{M}-1\right)+P_2^{\,2}\left(\frac{r}{M}-1\right)\right]\,.\label{eq:H extsol}
\end{align}
\end{widetext}
Similarly, $K$ can be obtained from $H_0$ via Eq.~\eqref{eq:Hsol}, i.e.,
\begin{equation}\label{eq:K extsol}
    K(r) = \frac{r^2-Mr-M^2}{r(r-2M)}H_0(r)+\frac{M}{2}H_0^\prime(r)\,.
\end{equation}
Substituting Eq.~\eqref{eq:H extsol} and Eq.~\eqref{eq:K extsol} into Eq.~\eqref{eq:zt}, we get the analytical solution of Detweiler redshift for $(\ell,m)=(2,0)$ mode. The result expanded to 7PN order is:
\begin{widetext}
\begin{equation}\label{eq:z20 7PN}
    z_{\rm SF,tid}^{20}(x) =-\frac{3\lambda x^6}{4M^5}\Big(1+7x+\frac{226}{7}x^2+\frac{872}{7}x^3+\frac{257003}{588}x^4+\frac{852697}{588}x^{5}+\frac{30056977}{6468}x^{6}+\frac{94191283}{6468}x^{7}\Big)\,.
\end{equation}
\end{widetext}

\section{Halmitonian formalism}\label{sec: appendix H}
In Refs.~\cite{Moncrief:1974am,Moncrief:1974an} Moncrief developed a Hamiltonian description for the perturbations of a perfect fluid star, both in the stellar interior and in the exterior Schwarzschild spacetime, 
\begin{equation}
    \mathscr{H} = \mathscr{H} _I +\mathscr{H}_E\,.
\end{equation}
For the fluid interior, the Hamiltonian is given by 
\begin{widetext}
\begin{equation}
\begin{split}
    16\pi \mathscr{H} _I = &\int_0^Rdr\Bigg\{e^{\nu-\lambda}\Bigg[\frac{\ell(\ell+1)}{2(\ell-1)(\ell+2)}\left(\Lambda \pi_1-8\pi r^2\varepsilon^\prime \pi_2\right)^2+8\pi e^{-2\lambda}r^2(p+\varepsilon)\left(\pi_1-\pi_{2}^{\prime}\right)^2\\
    &+8\pi r(p+\varepsilon)\ell(\ell+1)\pi_1\pi_2+8\pi(p+\varepsilon)\ell(\ell+1)\pi_2^2
    \Bigg]\Bigg\}\\
    &+\int_0^Rdr\Bigg\{\frac{e^{\lambda+\nu}}{\Lambda^2}\Bigg[\frac{(\ell-1)(\ell+2)}{2\ell(\ell+1)}e^{-2\lambda}\left(q_1-q_{2}^\prime\right)^2-\frac{(\ell-1)(\ell+2)}{2r}q_1q_2\\
    &+\frac{(\ell-1)(\ell+2)}{2r^2\Lambda}q_1^2\left[(\ell-1)(\ell+2)-8\pi r^3\varepsilon^\prime\right]+\frac{\partial p/\partial\varepsilon}{32\pi r^2(p+\varepsilon)}\left(\Lambda q_2+8\pi r^2\varepsilon^\prime q_1\right)^2\Bigg]\Bigg\}\\
    &+\int_0^Rdr\Bigg\{e^{\lambda+\nu}\Bigg[\frac{\pi_4^2}{2\ell(\ell+1)}-2e^{-2\lambda}\pi_4\left(\pi_1-\pi_{2}^\prime\right)-\frac{\ell(\ell+1)-\Lambda}{r}\pi_2\pi_4\\
    &+\frac{2e^{-2\lambda}}{(\ell-1)\ell(\ell+1)(\ell+2)r^2}\pi_3^2+\frac{2e^{-2\lambda}}{(\ell-1)(\ell+2)r^2}\pi_3\left(r\Lambda\pi_1+\pi_2(r\Lambda)^\prime\right)\Bigg]\Bigg\}\\
    &+\int_0^Rdr\Bigg\{\frac{e^{\nu-\lambda}}{8\pi r^2(p+\varepsilon)}\Bigg[\frac{1}{4}e^{2\lambda}\pi_4^2+\frac{r^2}{\ell(\ell+1)}\left[\frac{1}{2}\pi_{4}^\prime+\frac{\pi_4}{r}-\frac{\pi_3}{r^2}\right]^2\Bigg]\Bigg\}\,,
\end{split}
\end{equation}
\end{widetext}
The variables $q_1,...,q_4$ and their conjugate momentum $\pi_1,...,\pi_4$ are related to the metric perturbation functions as (in Regge-Wheeler gauge)
\begin{widetext}
\begin{align}
    q_1 =& \ell(\ell+1)rK-2e^{-2\lambda}r(K-H_0+rK^\prime+r\lambda^\prime K)\,,\\
    q_2 =&e^{-2\lambda}\Bigg\{\left[2+e^{2\lambda}\ell(\ell+1)-4r\lambda^\prime\right]H_0+2rH_{0}^\prime+2rK^\prime(-3+r\lambda^\prime)-2r^2K^{\prime\prime}\\
    &+\left[-2+e^{2\lambda}\ell(\ell+1)+4r^2\lambda^{\prime2}-2r^2\lambda^{\prime\prime}\right]K\Bigg\}\,,\\
    q_3 = &q_4 =0\,,
\end{align}
and 
\begin{align}
    e^{\lambda+\nu}\pi_4 = &-\ell(\ell+1)H_1+4r(\lambda^\prime+\nu^\prime)e^{-2\lambda}H_1+2r\left[r\dot K^\prime+(1-r\nu^\prime)\dot K-\dot H_0\right]\,,\\
    e^{\lambda+\nu}\pi_3 = &e^{-2\lambda}r\left(6r\lambda^\prime+6r\nu^\prime-6r^2\lambda^{\prime2}-2r^2\nu^{\prime2}-8r^2\lambda^\prime\nu^\prime+2r^2\lambda^{\prime\prime}+2r^2\nu^{\prime\prime}\right)H_1\nonumber\\
    &+\ell(\ell+1)(r\lambda^\prime-1)rH_1 -\ell(\ell+1)r^2H_{1}^\prime+2r^3(\lambda^\prime+\nu^\prime)e^{-2\lambda}H_{1}^\prime\nonumber\\
    &+\frac{\ell(\ell+1)}{2}e^{2\lambda}r^2(\dot H_0+\dot K)+(r\lambda^\prime+r\nu^\prime-3)r^2\dot H_0-(r\lambda^\prime+2r\nu^\prime-5)r^3\dot K^\prime\nonumber\\
    &+\left(3-5r\nu^\prime+r^2\nu^{\prime2}-r\lambda^\prime+r^2\lambda^\prime\nu^\prime-r^2\nu^{\prime\prime}\right)r^2\dot K-r^3\dot H_{0}^\prime+r^4\dot K^{\prime\prime}\,,\\
    e^{\lambda+\nu}\pi_2=&\frac{2rH_1}{\Lambda}- \frac{4r^2e^{-2\lambda}}{\ell(\ell+1)\Lambda}(\lambda^\prime+\nu^\prime)H_1+\frac{2r^2}{\ell(\ell+1)\Lambda}\dot H_0-\frac{2r^3}{\ell(\ell+1)\Lambda}\dot K^\prime\nonumber\\
    &+\frac{2r^2(r\nu^\prime-1)}{\ell(\ell+1)\Lambda}\dot K-\frac{r^2e^{2\lambda}}{\Lambda}\dot K\,,\\
    e^{\lambda+\nu}\pi_1 = &-\frac{2re^{-2\lambda}}{\ell(\ell+1)\Lambda^2}\big[(4\nu^\prime-6r\lambda^{\prime2}-2r\nu^{\prime2}+4\lambda^\prime-8r\lambda^\prime\nu^\prime+2r\lambda^{\prime\prime}+2r\nu^{\prime\prime})\Lambda\nonumber\\
    &+\ell(\ell+1)e^{2\lambda}(\Lambda^\prime+\Lambda\lambda^\prime)-2r(\lambda^\prime+\nu^\prime)\Lambda^\prime\big]H_1\nonumber\\
    &+\frac{2re^{-2\lambda}}{\ell(\ell+1)\Lambda}\left[\ell(\ell+1)e^{2\lambda}-2r(\lambda^\prime+\nu^\prime)\right]H_{1}^\prime\nonumber\\
    &-\frac{r}{\ell(\ell+1)\Lambda^2}\left[\ell(\ell+1)e^{2\lambda}\Lambda+2r(\lambda^\prime+\nu^\prime)\Lambda-4\Lambda+2r\Lambda^\prime\right]\dot H_0\nonumber\\
    &+\frac{r}{\ell(\ell+1)\Lambda^2}\big[(8r\nu^\prime-4-2r^2\nu^{\prime2}+2r\lambda^\prime-2r^2\lambda^\prime\nu^\prime+2r^2\nu^{\prime\prime})\Lambda\nonumber\\
    &+\ell(\ell+1)re^{2\lambda}\Lambda^\prime+(2r-2r^2\nu^\prime)\Lambda^\prime\big]\dot K\nonumber\\
    &+\frac{2r^2}{\ell(\ell+1)\Lambda}\dot H_{0}^\prime+\frac{2r^2}{\ell(\ell+1)\Lambda^2}\left[(2r\nu^\prime+r\lambda^\prime-4)\Lambda+r\Lambda^\prime\right]\dot K^\prime-\frac{2r^3}{\ell(\ell+1)\Lambda}\dot K^{\prime\prime}\,,
\end{align}
\end{widetext}
where 
\begin{equation}
    \Lambda = \ell(\ell+1)-2e^{-2\lambda}(1+r\lambda^\prime)\,,
\end{equation}
and an overdot denotes differentiation with respect to $t$.

For the external Schwarzschild spacetime, we find that in the frequency domain considered in this work, the original Hamiltonian in Ref.~\cite{Moncrief:1974am} can be greatly simplified. The resulting expression is
\begin{equation}
    \mathscr{H} _E =  \frac{1}{64\pi}\frac{(\ell+2)!}{(\ell-1)!}\left\{\int^\infty_{R_*} \omega^2 Z^2dr_* +\left[\frac{Z\Lambda (Z\Lambda)_{r_*}}{2\Lambda^2}\right]_{R_*}^\infty\right\}\,.
\end{equation}
Here $r_*$ is the tortoise coordinate and $Z$ is the Zerilli variable.

\section{An alternative way to compute ${\rm W}_n$}\label{sec:alter}
\subsection{Formalism}
Let us first consider Regge-wheeler/Zerilli equation outside the star which is given by
\begin{align}
\left ( \frac{\partial^2}{\partial r_*^{2}}+\omega^2-V(r)\right )\psi=0\,,
\end{align}
with matching condition at $r\to R$ and boundary condition at $r_* \to \infty$ as
\begin{align}
\psi \to A_{\rm in} e^{-i \omega r_*} + A_{\rm out} e^{i \omega r_*}\,.
\end{align} 
Suppose for some particular $\omega_0$ the ingoing wave is zero: $A_{\rm in}(\omega_0) =0$. So what is $d A_{\rm in}/d \omega (\omega_0)$ in this case? To answer this question, we first define $\psi_1 = \psi$, $\psi_2=\partial\psi/\partial r_*$ and $\Vec{\psi}=(\psi_1, \psi_2)$, then we have
\begin{equation}
\left\{
             \begin{array}{lr}
             \frac{\partial\psi_{1}}{\partial r_*}-\psi_{2}=0&  \\
             \frac{\partial\psi_2}{\partial r_*}+(\omega^2-V)\psi_1=0\,, & 
             \end{array}
\right.
\end{equation}
These equations can also be written as
\begin{align}\label{eq:mathcal_O_1}
    \begin{pmatrix} \frac{\partial}{\partial r_*} & -1 \\ \omega^2-V & \frac{\partial}{\partial r_*}  \\ \end{pmatrix} 
\begin{pmatrix} \psi_1 \\ \psi_2  \\ \end{pmatrix}
=\mathcal{O}\Vec{\psi}=0\,,
\end{align}
and the boundary condition at infinity $r_*\to\infty$ is
\begin{align}
    \psi_1&\to A_{\rm in}e^{-i\omega r_*}+A_{\rm out}e^{i\omega r_*}\,,\\
    \psi_2&\to -i\omega A_{\rm in}e^{-i\omega r_*}+i\omega A_{\rm out}e^{i\omega r_*}\,.
\end{align}
In order to compute $d A_{\rm in}/d \omega (\omega_0)$, we define $\phi_1=e^{-i\omega r_*}\psi_1$,  and $\phi_2=\partial\phi_1/\partial {r_*}$, then we can rewrite Eq.~\eqref{eq:mathcal_O_1} as
\begin{align}\label{eq:mathcal_O_2}
    \begin{pmatrix} \frac{\partial}{\partial r_*} & -1 \\ -V & \frac{\partial}{\partial r_*}+2i\omega  \\ \end{pmatrix} 
\begin{pmatrix} \phi_1 \\ \phi_2  \\ \end{pmatrix}
=\mathcal{\hat{O}}\Vec{\phi}=0\,.
\end{align}
At infinity $r\to\infty$, $\Vec{\phi}$ behave as
\begin{align}
    \phi_1&\to A_{\rm in}e^{-2i\omega r_*}+A_{\rm out}\,,\\
    \phi_2&\to -2i\omega A_{\rm in} e^{-2i\omega r_*}\,.
\end{align}
Now since $\mathcal{\hat{O}}(\omega_0) \Vec{\phi}(\omega_0) =0$, at $\omega=\omega_0+\delta \omega$ we have $\Vec{\phi}(\omega)=\Vec{\phi}(\omega_0)+\delta \Vec{\phi}$ and
\begin{align}\label{eq:;pert1}
\mathcal{\hat{O}}(\omega_0) \delta \Vec{\phi} +\delta \omega\frac{\partial \mathcal{\hat{O}}}{\partial \omega}(\omega_0)\Vec{\phi}(\omega_0)=0\,,
\end{align}
where 
\begin{align}
    \frac{\partial \mathcal{\hat{O}}}{\partial \omega}(\omega_0) = \begin{pmatrix} 0 & 0 \\ 0 & 2i  \\ \end{pmatrix}\,.
\end{align}

Here $\delta \phi$ can be solved by imposing the ingoing boundary condition 
\begin{widetext}
    \begin{align}
    \frac{\delta\phi_1}{\delta\omega}\to&\frac{\delta A_{\rm in}}{\delta\omega}e^{-2i\omega r_*}\left(\bar{a}_0+\frac{\bar{a}_1}{r}+\frac{\bar{a}_2}{r^2}\right)+A_{\rm in}(-2ir_*)e^{-2i\omega r_*}\left(\bar{a}_0+\frac{\bar{a}_1}{r}+\frac{\bar{a}_2}{r^2}\right)\nonumber\\
    &+A_{\rm in}e^{-2i\omega r_*}\left(\bar{a}_0+\frac{\bar{a}_1}{r}+\frac{\bar{a}_2}{r^2}\right)_{,\omega}+\frac{\delta A_{\rm out}}{\delta\omega}\left(a_0+\frac{a_1}{r}+\frac{a_2}{r^2}\right)+A_{\rm out}\left(a_0+\frac{a_1}{r}+\frac{a_2}{r^2}\right)_{,\omega}\,,\\
    \frac{\delta\phi_2}{\delta\omega}\to&\frac{\delta A_{\rm in}}{\delta\omega}(-2i\omega)e^{-2i\omega r_*}\left(\bar{a}_0+\frac{\bar{a}_1}{r}+\frac{\bar{a}_2}{r^2}\right)+A_{\rm in}(-2i)e^{-2i\omega r_*}\left(\bar{a}_0+\frac{\bar{a}_1}{r}+\frac{\bar{a}_2}{r^2}\right)\nonumber\\
    &+A_{\rm in}(-2i\omega)(-2ir_*)e^{-2i\omega r_*}\left(\bar{a}_0+\frac{\bar{a}_1}{r}+\frac{\bar{a}_2}{r^2}\right)+A_{\rm in}(-2i\omega)e^{-2i\omega r_*}\left(\bar{a}_0+\frac{\bar{a}_1}{r}+\frac{\bar{a}_2}{r^2}\right)_{,\omega}\nonumber\\
    &+\frac{\delta A_{\rm in}}{\delta\omega}e^{-2i\omega r_*}\left(-\frac{\bar{a}_1}{r^2}-2\frac{\bar{a}_2}{r^3}\right)\frac{dr}{dr_*}+A_{\rm in}(-2ir_*)e^{-2i\omega r_*}\left(-\frac{\bar{a}_1}{r^2}-2\frac{\bar{a}_2}{r^3}\right)\frac{dr}{dr_*}\nonumber\\
    &+A_{\rm in}e^{-2i\omega r_*}\left(-\frac{\bar{a}_1}{r^2}-2\frac{\bar{a}_2}{r^3}\right)_{,\omega}\frac{dr}{dr_*}+\frac{\delta A_{\rm out}}{\delta\omega}\left(-\frac{a_1}{r^2}-2\frac{a_2}{r^3}\right)\frac{dr}{dr_*}+A_{\rm out}\left(-\frac{a_1}{r^2}-2\frac{a_2}{r^3}\right)_{,\omega}\frac{dr}{dr_*}\,,
\end{align}
\end{widetext}
at $r \to \infty$. If $G(x,x')$ is the corresponding Green's function, then
\begin{align}
\delta \phi =\delta \omega \int^\infty_{-\infty} d x' G(x,x') \frac{\partial \mathcal{\hat{O}}}{\partial \omega} \phi_0(x')\,,
\end{align}
and
\begin{align}
\frac{d A_{\rm in}}{d \omega}(\omega_0)=e^{2 i \omega x} \int^\infty_{-\infty} d x' G(x,x') \frac{\partial \mathcal{\hat{O}}}{\partial \omega} \phi_0(x'), \quad x\to \infty\,.
\end{align}
For the relativistic star case, we may accordingly redefine the Zerilli master variable to be $Z e^{-i \omega r_*}$ and use the modified wave equation outside the star. Accordingly we expect
\begin{align}
\frac{d A_{\rm in}}{d \omega}(\omega_0)=e^{2 i \omega r_*} \int^\infty_{0} d x' G(r_*,x') \frac{\partial \mathcal{L}}{\partial \omega} \phi_0(x')\,, \quad r_*\to \infty\,.
\end{align}

We can rewrite interior equations \eqref{eq:H1}, \eqref{eq:K}, \eqref{eq:W} and \eqref{eq:X} as follows:
\begin{align}\label{eq:delta_equation_interior}
    \frac{d\Vec{u}}{dr}-\Vec{M}(r, \omega)\Vec{u}(r):=\mathcal{P}\Vec{u}=0\,.
\end{align}
Then similar to the above discussion, at $\omega=\omega_0+\delta \omega$ we have $\Vec{u}=\Vec{u}_0+\delta \Vec{u}$ with $\Vec{u}_0=\{H_1^{0},K^{0},W^{0},X^{0}\}$ a solution of the interior equations, and then we have
\begin{align}\label{eq:;pert2}
\mathcal{P}(\omega_0) \delta \Vec{u} +\delta \omega\frac{\partial \mathcal{P}}{\partial \omega} \Vec{u}_0=0\,,
\end{align}
where
\begin{widetext}
    \begin{align}
    \frac{\partial \mathcal{P}}{\partial \omega}=\Bigg\{
    \frac{e^{\lambda}}{r}\left[\frac{\delta H_0}{\delta\omega}-16\pi(\varepsilon+p)\frac{\delta V}{\delta\omega}\right], \frac{1}{r}\frac{\delta H_0}{\delta\omega}, re^{\lambda/2}\left[-\frac{\ell(\ell+1)}{r^2}\frac{\delta V}{\delta\omega}+\frac{1}{2}\frac{\delta H_0}{\delta\omega}\right] , Q
    \Bigg\}\,,
\end{align}
\end{widetext}
with
\begin{align}
    Q=&(p+\varepsilon)e^{\nu/2}\bigg[\left(\frac{\nu^\prime}{2}-\frac{1}{2r}\right)\frac{\delta H_0}{\delta\omega}-\frac{\ell(\ell+1)}{2r^2}\nu^\prime \frac{\delta V}{\delta\omega}\nonumber\\
    &-\frac{2\omega}{r}e^{\lambda/2-\nu}W^0\bigg]\,,\nonumber
\end{align}
in which $\frac{\delta H_0}{\delta\omega}$ and $\frac{\delta V}{\delta\omega}$ are given by
\begin{align}
    \frac{\delta H_0}{\delta\omega}=&\frac{1}{hr}\left(\frac{2\omega}{e^{\lambda+\nu}}H_1^0-\frac{2\omega}{e^{\lambda}}K^0\right)\,,\\
    \frac{\delta V}{\delta\omega}=&-2\omega^{-3}\frac{e^{\nu/2}X^0}{\varepsilon+p}-2\omega^{-3}\frac{e^{\nu/2}}{\varepsilon+p}\frac{1}{r}e^{\lambda+\nu}W^0\,,\\
    &-\frac{1}{2}e^{\nu}(-2\omega^{-3})H_1^0-\frac{1}{2}e^{\nu}\omega^{-2}\frac{\delta H_0}{\delta\omega}\,,\nonumber\\
    hr=&3M(r)+\frac{1}{2}(\ell+2)(\ell-1)r+4\pi r^3 p(r)\,.
\end{align}
%-\frac{l(l+1)}{r^2}\frac{\delta V}{\delta\omega}
The boundary conditions of the above equations near the center $r=0$,
\begin{align}
    \delta \Vec{u}|_{\rm near\ center}=\frac{d\Vec{u}}{d\omega}\cdot \delta\omega_{\rm near\ center}\,.
\end{align}
As a result, we  assume that the solutions are represented by power series expansions near the center $r=0$, which are given by
\begin{align}\label{eq:zero}
	\frac{\delta H_{1}}{\delta\omega}&=\frac{\delta y_{0}}{\delta\omega}+\frac{1}{2}\frac{\delta y_{2}}{\delta\omega}r^2+\cdots,\nonumber\\
	\frac{\delta K}{\delta\omega}&=\frac{\delta k_0}{\delta\omega}+\frac{1}{2}\frac{\delta k_2}{\delta\omega}r^2+\cdots,\nonumber\\
	\frac{\delta W}{\delta\omega}&=\frac{\delta w_0}{\delta\omega}+\frac{1}{2}\frac{\delta w_2}{\delta\omega}r^2+\cdots,\nonumber\\
	\frac{\delta X}{\delta\omega}&=\frac{\delta x_0}{\delta\omega}+\frac{1}{2}\frac{\delta x_2}{\delta\omega}r^2+\cdots.
\end{align} 
If we substitute Eqs.~\eqref{eq:zero} into Eqs.~\eqref{eq:delta_equation_interior} and solve the equations order by order, we can obtain the first order constraints $\mathcal{O}(r^0)$ and the second order constraints $\mathcal{O}(r^2)$. In particular,  the first order constraints $\mathcal{O}(r^0)$, following these relations can be written as
\begin{align}
	\frac{\delta x_0}{\delta\omega}&=(\varepsilon_0+p_0)e^{\nu_0/2}\biggl\{\bigg[\frac{4\pi}{3}(\varepsilon_0+3p_0)-\omega^2e^{-\nu_0}/\ell\bigg]\frac{\delta w_0}{\delta\omega}\nonumber\\
    &+\frac{1}{2}\frac{\delta k_0}{\delta\omega}\biggr\}+(\varepsilon_0+p_0)e^{\nu_0/2}(-2\omega e^{-\nu_0}/\ell)W_{r0}\,,\label{eq:regular11}\\
	\frac{\delta y_0}{\delta\omega}&=\frac{2\ell \frac{\delta k_0}{\delta\omega}+16\pi(\varepsilon_0+p_0)\frac{\delta w_0}{\delta\omega}}{\ell(\ell+1)}\,.\label{eq:regular12}
\end{align}
Let $\Vec{u}_{2}=\{y_2, k_2, w_2, x_2,\}$, then we can rewrite the Eqs.~(36)-(39) in~\cite{Feng:2024olt} as
\begin{align}
    \Vec{F}(\omega)\Vec{u}_2=0\,.
\end{align}
By implicit function theorem, $\frac{\delta y_2}{\delta\omega}$, $\frac{\delta k_2}{\delta\omega}$,$\frac{\delta w_2} {\delta\omega}$ and $\frac{\delta x_2}{\delta\omega}$ can be obtained.

For the even parity, the Zerilli function $Z$ is related to interior perturbation functions at surface as
\begin{align}
	Z(r_*)&=-\frac{r^2e^{-\lambda}}{nr+3M}H_1+\frac{r^2}{nr+3M}K\,,\label{eq:boundary21}\\
	\frac{dZ(r_*)}{dr_*}&=\frac{n(n+1)r^2+3nMr+6M^2}{(nr+3M)^2}e^{-\lambda}H_{1}\nonumber\\
 &\quad-\frac{nr^2-3nMr-3M^2}{(nr+3M)^2}K\,,\label{eq:boundary22}
\end{align}
which is evaluated at $r=R$. This can be rewritten as the form
\begin{align}
    \begin{pmatrix} Z  \\ \frac{dZ(r_*)}{dr_*}  \\ \end{pmatrix} = \begin{pmatrix} A_{11} & A_{12} \\ A_{21} & A_{22}  \\ \end{pmatrix}
    \begin{pmatrix} H_1  \\ K  \\ \end{pmatrix}\,.
\end{align}
Then the relationship between $\delta\Vec{\psi}$ and $\delta H_1$, $\delta K$ are
\begin{align}
    \delta\psi_1&=A_{11}\delta H_1+A_{12}\delta K+\frac{\partial A_{11}}{\partial\omega}H_{1}(\omega_0)\delta\omega\nonumber\\
    &\quad+\frac{\partial A_{12}}{\partial\omega}K(\omega_0)\delta\omega\,,\\
    \delta\psi_2&=A_{21}\delta H_1+A_{22}\delta K+\frac{\partial A_{21}}{\partial\omega}H_{1}(\omega_0)\delta\omega\nonumber\\
    &\quad+\frac{\partial A_{22}}{\partial\omega}K(\omega_0)\delta\omega\,.
\end{align}

\subsection{Shooting Method}
Eqs.~(\ref{eq:;pert1} \& \ref{eq:;pert2}) are nonhomogeneous differential equations.
The boundary conditions include two regularity conditions at the center of the star, one condition $\delta X/\delta\omega$ = 0 imposed at the stellar surface, and another condition $\delta A_{\rm out}/\delta\omega = 0$ applied at spatial infinity. The shooting method, which transforms a boundary value problem into an initial value problem, can be employed to solve this system effectively.

We begin by making an initial guess for 
$\delta\Vec{u}/\delta\omega(r_0)$ at the center and $\delta\Vec{u}/\delta\omega(R)$ at the surface of the star. These two solutions must match at 
$r=R/2$. From there, we adjust our guess iteratively, refining the solution until the boundary conditions at both the center and the surface of the star are satisfied. Additionally, we ensure that $\delta A_{\rm out}/\delta\omega=0$ is correctly determined at infinity.

\subsection{Results}
In Table~\ref{tab:Wn}, we list a set of numerical results for ${\rm W}_n$ (of $f$-modes) computed using the eigenfunction method, and compare them with the results obtained via the contour integral method described in Appendix~\ref{sec:appendix residue}. In our calculations, we adopt a polytropic equation of state with $n = 1$ and fix the stellar mass $M=1.4\,M_\odot$, while vary the stellar radius $R$. It can be seen that the results obtained from the eigenfunction method agree remarkably well with those from the contour integral method, with typical relative differences on the order of $10^{-3}$. The minor discrepancies observed are mainly attributed to variations in the numerical computation of the $f$-mode frequency.

{\renewcommand{\arraystretch}{1.2}
\begin{table*}[hbt!]
\begin{tabular*}{\textwidth}{ @{\extracolsep{\fill}} cccc|ccc}
\hline\hline
\multicolumn{1}{c}{\multirow{2}{*}{$M/R$}}
	&\multicolumn{3}{c}{{${\rm Re}({\rm W}_n)$}}
	&\multicolumn{3}{c}{{${\rm Im}({\rm W}_n)$}}\\
    \cline{2-7}
     &Cont. Int. &Eigenfunc.&Rel. Diff.&Cont. Int. &Eigenfunc.
&Rel. Diff.\\
    \hline
    $0.1736$ & $3.00578\times10^{-5}$ & $3.01606\times10^{-5}$ &0.00342 & $4.32034\times10^{-5}$ &$4.31338\times10^{-5}$ &0.00161 \\
    $0.1491$ & $1.82786\times10^{-5}$ & $1.83621\times10^{-5}$ &0.00456 & $2.92974\times10^{-5}$ &$2.92457\times10^{-5}$ &0.00177 \\
    $0.1202$ & $8.13349\times10^{-6}$ & $8.19500\times10^{-6}$ &0.00756 & $1.57176\times10^{-5}$ &$1.56856\times10^{-5}$ &0.00204 \\
    $0.1016$ & $4.08830\times10^{-6}$ & $4.07389\times10^{-6}$ &0.00352 & $9.28618\times10^{-6}$ &$9.31389\times10^{-6}$ &0.00298 \\
    $0.0799$ & $1.43370\times10^{-6}$ & $1.43332\times10^{-6}$ &0.00027 & $4.16154\times10^{-6}$ &$4.16166\times10^{-6}$ &0.00003 \\
    $0.0502$ & $1.65314\times10^{-7}$ & $1.65194\times10^{-7}$ &0.00073 & $7.93831\times10^{-7}$ &$7.93855\times10^{-7}$ &0.00003\\
    $0.0304$ & $1.41535\times10^{-8}$ & $1.40133\times10^{-8}$ &0.01000 & $1.20746\times10^{-8}$ &$1.20776\times10^{-8}$ &0.00025 \\
    $0.0185$ & $1.15318\times10^{-9}$ & $1.12708\times10^{-9}$ &0.02315 & $1.77977\times10^{-8}$ &$1.78004\times10^{-8}$ &0.00015 \\
    $0.0103$ & $5.62403\times10^{-11}$ & $5.69662\times10^{-11}$ &0.01274 & $1.79181\times10^{-9}$ &$1.79306\times10^{-9}$ &0.00070 \\
    $0.0062$ & $3.88375\times10^{-12}$ & $3.22998\times10^{-12}$ &0.20241 & $2.37554\times10^{-10}$ &$2.37565\times10^{-10}$ &0.00005 \\
\hline\hline
\end{tabular*}
\caption{The values of ${\rm W}_n$ (for $f$-modes) as a function of the compactness of the neutron star, computed using two independent numerical methods: the contour integral method (Appendix~\ref{sec:appendix residue}) and the eigenfunction method (Appendix~\ref{sec:alter}). The stellar models are constructed using a polytropic equation of state with index $n = 1$, with the mass fixed at $M = 1.4\,M_\odot$ and the radius $R$ varied over a range of values. The results from the two methods show excellent agreement, with relative differences typically at the level of $10^{-3}$.}\label{tab:Wn}
\end{table*}
}

%%%%%%%%%%%%%%%%%%%%%%%%%%%%%%%%%%%%%%%%%%%%%%%%%%%%%%%%%%%%%%%%%%%%%%%%%%%%%%%
\def\bibsection{\section*{References}}
%%%%%%%%%%%%%%%%%%%%%%%%%%%%%%%%%%%%%%%%%%%%%%%%%%%%%%%%%%%%%%%%%%%%%%%%%%%%%%%
\bibliography{references}

@article{Lai:1993di,
    author = "Lai, Dong",
    title = "{Resonant oscillations and tidal heating in coalescing binary neutron stars}",
    eprint = "astro-ph/9404062",
    archivePrefix = "arXiv",
    reportNumber = "CRSR-1064",
    doi = "10.1093/mnras/270.3.611",
    journal = "Mon. Not. Roy. Astron. Soc.",
    volume = "270",
    pages = "611",
    year = "1994"
}

@article{Lai:2006pr,
    author = "Lai, Dong and Wu, Yanqin",
    title = "{Resonant Tidal Excitations of Inertial Modes in Coalescing Neutron Star Binaries}",
    eprint = "astro-ph/0604163",
    archivePrefix = "arXiv",
    doi = "10.1103/PhysRevD.74.024007",
    journal = "Phys. Rev. D",
    volume = "74",
    pages = "024007",
    year = "2006"
}

@article{Yu:2017cxe,
    author = "Yu, Hang and Weinberg, Nevin N.",
    title = "{Dynamical tides in coalescing superfluid neutron star binaries with hyperon cores and their detectability with third generation gravitational-wave detectors}",
    eprint = "1705.04700",
    archivePrefix = "arXiv",
    primaryClass = "astro-ph.HE",
    doi = "10.1093/mnras/stx1188",
    journal = "Mon. Not. Roy. Astron. Soc.",
    volume = "470",
    number = "1",
    pages = "350--360",
    year = "2017"
}

@article{Pan:2020tht,
    author = "Pan, Zhen and Lyu, Zhenwei and Bonga, B\'eatrice and Ortiz, N\'estor and Yang, Huan",
    title = "{Probing Crust Meltdown in Inspiraling Binary Neutron Stars}",
    eprint = "2003.03330",
    archivePrefix = "arXiv",
    primaryClass = "astro-ph.HE",
    doi = "10.1103/PhysRevLett.125.201102",
    journal = "Phys. Rev. Lett.",
    volume = "125",
    number = "20",
    pages = "201102",
    year = "2020"
}

@article{Poisson:2020eki,
    author = "Poisson, Eric",
    title = "{Gravitomagnetic tidal resonance in neutron-star binary inspirals}",
    eprint = "2003.10427",
    archivePrefix = "arXiv",
    primaryClass = "gr-qc",
    doi = "10.1103/PhysRevD.101.104028",
    journal = "Phys. Rev. D",
    volume = "101",
    number = "10",
    pages = "104028",
    year = "2020"
}

@article{Kwon:2024zyg,
    author = "Kwon, K. J. and Yu, Hang and Venumadhav, Tejaswi",
    title = "{Resonance Locking of Anharmonic $g$-Modes in Coalescing Neutron Star Binaries}",
    eprint = "2410.03831",
    archivePrefix = "arXiv",
    primaryClass = "gr-qc",
    month = "10",
    year = "2024"
}

@article{Ma:2020oni,
    author = "Ma, Sizheng and Yu, Hang and Chen, Yanbei",
    title = "{Detecting resonant tidal excitations of Rossby modes in coalescing neutron-star binaries with third-generation gravitational-wave detectors}",
    eprint = "2010.03066",
    archivePrefix = "arXiv",
    primaryClass = "gr-qc",
    doi = "10.1103/PhysRevD.103.063020",
    journal = "Phys. Rev. D",
    volume = "103",
    number = "6",
    pages = "063020",
    year = "2021"
}

@article{Feng:2024olt,
    author = "Feng, Xuefeng and Yang, Huan",
    title = "{Universal gravitational self-force for a point mass orbiting around a compact star}",
    eprint = "2406.02101",
    archivePrefix = "arXiv",
    primaryClass = "gr-qc",
    doi = "10.1103/PhysRevD.110.044066",
    journal = "Phys. Rev. D",
    volume = "110",
    number = "4",
    pages = "044066",
    year = "2024"
}

@article{Yang:2017xlf,
    author = "Yang, Huan and Paschalidis, Vasileios and Yagi, Kent and Lehner, Luis and Pretorius, Frans and Yunes, Nicol\'as",
    title = "{Gravitational wave spectroscopy of binary neutron star merger remnants with mode stacking}",
    eprint = "1707.00207",
    archivePrefix = "arXiv",
    primaryClass = "gr-qc",
    doi = "10.1103/PhysRevD.97.024049",
    journal = "Phys. Rev. D",
    volume = "97",
    number = "2",
    pages = "024049",
    year = "2018"
}

@article{Yang:2018bzx,
    author = "Yang, Huan and East, William E. and Paschalidis, Vasileios and Pretorius, Frans and Mendes, Raissa F. P.",
    title = "{Evolution of Highly Eccentric Binary Neutron Stars Including Tidal Effects}",
    eprint = "1806.00158",
    archivePrefix = "arXiv",
    primaryClass = "gr-qc",
    doi = "10.1103/PhysRevD.98.044007",
    journal = "Phys. Rev. D",
    volume = "98",
    number = "4",
    pages = "044007",
    year = "2018"
}

@article{Yang:2019kmf,
    author = "Yang, Huan",
    title = "{Inspiralling eccentric binary neutron stars: Orbital motion and tidal resonance}",
    eprint = "1904.11089",
    archivePrefix = "arXiv",
    primaryClass = "gr-qc",
    doi = "10.1103/PhysRevD.100.064023",
    journal = "Phys. Rev. D",
    volume = "100",
    number = "6",
    pages = "064023",
    year = "2019"
}

@article{Zhang:2022yab,
    author = "Zhang, Teng and Yang, Huan and Martynov, Denis and Schmidt, Patricia and Miao, Haixing",
    title = "{Gravitational-Wave Detector for Postmerger Neutron Stars: Beyond the Quantum Loss Limit of the Fabry-Perot-Michelson Interferometer}",
    eprint = "2212.12144",
    archivePrefix = "arXiv",
    primaryClass = "gr-qc",
    doi = "10.1103/PhysRevX.13.021019",
    journal = "Phys. Rev. X",
    volume = "13",
    number = "2",
    pages = "021019",
    year = "2023"
}

@article{Ma:2024qcv,
    author = "Ma, Sizheng and Yang, Huan",
    title = "{Excitation of quadratic quasinormal modes for Kerr black holes}",
    eprint = "2401.15516",
    archivePrefix = "arXiv",
    primaryClass = "gr-qc",
    doi = "10.1103/PhysRevD.109.104070",
    journal = "Phys. Rev. D",
    volume = "109",
    number = "10",
    pages = "104070",
    year = "2024"
}

@article{Martynov:2019gvu,
    author = "Martynov, Denis and others",
    title = "{Exploring the sensitivity of gravitational wave detectors to neutron star physics}",
    eprint = "1901.03885",
    archivePrefix = "arXiv",
    primaryClass = "astro-ph.IM",
    doi = "10.1103/PhysRevD.99.102004",
    journal = "Phys. Rev. D",
    volume = "99",
    number = "10",
    pages = "102004",
    year = "2019"
}

@article{Miao:2017qot,
    author = "Miao, Haixing and Yang, Huan and Martynov, Denis",
    title = "{Towards the Design of Gravitational-Wave Detectors for Probing Neutron-Star Physics}",
    eprint = "1712.07345",
    archivePrefix = "arXiv",
    primaryClass = "gr-qc",
    doi = "10.1103/PhysRevD.98.044044",
    journal = "Phys. Rev. D",
    volume = "98",
    number = "4",
    pages = "044044",
    year = "2018"
}

@article{Steinhoff:2016rfi,
    author = "Steinhoff, Jan and Hinderer, Tanja and Buonanno, Alessandra and Taracchini, Andrea",
    title = "{Dynamical Tides in General Relativity: Effective Action and Effective-One-Body Hamiltonian}",
    eprint = "1608.01907",
    archivePrefix = "arXiv",
    primaryClass = "gr-qc",
    doi = "10.1103/PhysRevD.94.104028",
    journal = "Phys. Rev. D",
    volume = "94",
    number = "10",
    pages = "104028",
    year = "2016"
}

@article{Leaver:1986gd,
    author = "Leaver, Edward W.",
    title = "{Spectral decomposition of the perturbation response of the Schwarzschild geometry}",
    doi = "10.1103/PhysRevD.34.384",
    journal = "Phys. Rev. D",
    volume = "34",
    pages = "384--408",
    year = "1986"
}

@article{Onozawa:1996ux,
    author = "Onozawa, Hisashi",
    title = "{A Detailed study of quasinormal frequencies of the Kerr black hole}",
    eprint = "gr-qc/9610048",
    archivePrefix = "arXiv",
    reportNumber = "TIT-HEP-344, COSMO-78",
    doi = "10.1103/PhysRevD.55.3593",
    journal = "Phys. Rev. D",
    volume = "55",
    pages = "3593--3602",
    year = "1997"
}

@article{Cook:2014cta,
    author = "Cook, Gregory B. and Zalutskiy, Maxim",
    title = "{Gravitational perturbations of the Kerr geometry: High-accuracy study}",
    eprint = "1410.7698",
    archivePrefix = "arXiv",
    primaryClass = "gr-qc",
    doi = "10.1103/PhysRevD.90.124021",
    journal = "Phys. Rev. D",
    volume = "90",
    number = "12",
    pages = "124021",
    year = "2014"
}

@article{Berti:2003jh,
    author = "Berti, Emanuele and Cardoso, Vitor and Kokkotas, Kostas D. and Onozawa, Hisashi",
    title = "{Highly damped quasinormal modes of Kerr black holes}",
    eprint = "hep-th/0307013",
    archivePrefix = "arXiv",
    doi = "10.1103/PhysRevD.68.124018",
    journal = "Phys. Rev. D",
    volume = "68",
    pages = "124018",
    year = "2003"
}

@article{Motohashi:2024fwt,
    author = "Motohashi, Hayato",
    title = "{Resonant Excitation of Quasinormal Modes of Black Holes}",
    eprint = "2407.15191",
    archivePrefix = "arXiv",
    primaryClass = "gr-qc",
    doi = "10.1103/PhysRevLett.134.141401",
    journal = "Phys. Rev. Lett.",
    volume = "134",
    number = "14",
    pages = "141401",
    year = "2025"
}

@article{Pitre:2023xsr,
    author = "Pitre, Tristan and Poisson, Eric",
    title = "{General relativistic dynamical tides in binary inspirals without modes}",
    eprint = "2311.04075",
    archivePrefix = "arXiv",
    primaryClass = "gr-qc",
    doi = "10.1103/PhysRevD.109.064004",
    journal = "Phys. Rev. D",
    volume = "109",
    number = "6",
    pages = "064004",
    year = "2024"
}

@article{Flanagan:2007ix,
    author = "Flanagan, Eanna E. and Hinderer, Tanja",
    title = "{Constraining neutron star tidal Love numbers with gravitational wave detectors}",
    eprint = "0709.1915",
    archivePrefix = "arXiv",
    primaryClass = "astro-ph",
    doi = "10.1103/PhysRevD.77.021502",
    journal = "Phys. Rev. D",
    volume = "77",
    pages = "021502",
    year = "2008"
}

@article{Kwon:2025zbc,
    author = "Kwon, K. J. and Yu, Hang and Venumadhav, Tejaswi",
    title = "{Resonance locking: radian-level phase shifts due to nonlinear hydrodynamics of $g$-modes in merging neutron star binaries}",
    eprint = "2503.11837",
    archivePrefix = "arXiv",
    primaryClass = "gr-qc",
    month = "3",
    year = "2025"
}

@article{Lau:2020bfq,
    author = "Lau, Shu Yan and Yagi, Kent",
    title = "{Probing hybrid stars with gravitational waves via interfacial modes}",
    eprint = "2012.13000",
    archivePrefix = "arXiv",
    primaryClass = "astro-ph.HE",
    doi = "10.1103/PhysRevD.103.063015",
    journal = "Phys. Rev. D",
    volume = "103",
    number = "6",
    pages = "063015",
    year = "2021"
}

@article{Hinderer:2007mb,
    author = "Hinderer, Tanja",
    title = "{Tidal Love numbers of neutron stars}",
    eprint = "0711.2420",
    archivePrefix = "arXiv",
    primaryClass = "astro-ph",
    doi = "10.1086/533487",
    journal = "Astrophys. J.",
    volume = "677",
    pages = "1216--1220",
    year = "2008",
    note = "[Erratum: Astrophys.J. 697, 964 (2009)]"
}

@article{Hinderer:2016eia,
    author = "Hinderer, Tanja and others",
    title = "{Effects of neutron-star dynamic tides on gravitational waveforms within the effective-one-body approach}",
    eprint = "1602.00599",
    archivePrefix = "arXiv",
    primaryClass = "gr-qc",
    doi = "10.1103/PhysRevLett.116.181101",
    journal = "Phys. Rev. Lett.",
    volume = "116",
    number = "18",
    pages = "181101",
    year = "2016"
}

@article{LIGOScientific:2017vwq,
    author = "Abbott, B. P. and others",
    collaboration = "LIGO Scientific, Virgo",
    title = "{GW170817: Observation of Gravitational Waves from a Binary Neutron Star Inspiral}",
    eprint = "1710.05832",
    archivePrefix = "arXiv",
    primaryClass = "gr-qc",
    reportNumber = "LIGO-P170817",
    doi = "10.1103/PhysRevLett.119.161101",
    journal = "Phys. Rev. Lett.",
    volume = "119",
    number = "16",
    pages = "161101",
    year = "2017"
}

@article{Shibata:2006nm,
    author = "Shibata, Masaru and Taniguchi, Keisuke",
    title = "{Merger of binary neutron stars to a black hole: disk mass, short gamma-ray bursts, and quasinormal mode ringing}",
    eprint = "astro-ph/0603145",
    archivePrefix = "arXiv",
    doi = "10.1103/PhysRevD.73.064027",
    journal = "Phys. Rev. D",
    volume = "73",
    pages = "064027",
    year = "2006"
}

@article{Kiuchi:2009jt,
    author = "Kiuchi, Kenta and Sekiguchi, Yuichiro and Shibata, Masaru and Taniguchi, Keisuke",
    title = "{Longterm general relativistic simulation of binary neutron stars collapsing to a black hole}",
    eprint = "0904.4551",
    archivePrefix = "arXiv",
    primaryClass = "gr-qc",
    doi = "10.1103/PhysRevD.80.064037",
    journal = "Phys. Rev. D",
    volume = "80",
    pages = "064037",
    year = "2009"
}

@article{Bauswein:2011tp,
    author = "Bauswein, A. and Janka, H. -Th.",
    title = "{Measuring neutron-star properties via gravitational waves from binary mergers}",
    eprint = "1106.1616",
    archivePrefix = "arXiv",
    primaryClass = "astro-ph.SR",
    doi = "10.1103/PhysRevLett.108.011101",
    journal = "Phys. Rev. Lett.",
    volume = "108",
    pages = "011101",
    year = "2012"
}

@article{Yang:2014tla,
    author = "Yang, Huan and Zimmerman, Aaron and Lehner, Luis",
    title = "{Turbulent Black Holes}",
    eprint = "1402.4859",
    archivePrefix = "arXiv",
    primaryClass = "gr-qc",
    doi = "10.1103/PhysRevLett.114.081101",
    journal = "Phys. Rev. Lett.",
    volume = "114",
    pages = "081101",
    year = "2015"
}

@article{Bauswein:2015vxa,
    author = "Bauswein, Andreas and Stergioulas, Nikolaos and Janka, Hans-Thomas",
    title = "{Exploring properties of high-density matter through remnants of neutron-star mergers}",
    eprint = "1508.05493",
    archivePrefix = "arXiv",
    primaryClass = "astro-ph.HE",
    doi = "10.1140/epja/i2016-16056-7",
    journal = "Eur. Phys. J. A",
    volume = "52",
    number = "3",
    pages = "56",
    year = "2016"
}

@article{Palenzuela:2015dqa,
    author = "Palenzuela, Carlos and Liebling, Steven L. and Neilsen, David and Lehner, Luis and Caballero, O. L. and O'Connor, Evan and Anderson, Matthew",
    title = "{Effects of the microphysical Equation of State in the mergers of magnetized Neutron Stars With Neutrino Cooling}",
    eprint = "1505.01607",
    archivePrefix = "arXiv",
    primaryClass = "gr-qc",
    doi = "10.1103/PhysRevD.92.044045",
    journal = "Phys. Rev. D",
    volume = "92",
    number = "4",
    pages = "044045",
    year = "2015"
}

@article{Baiotti:2016qnr,
    author = "Baiotti, Luca and Rezzolla, Luciano",
    title = "{Binary neutron star mergers: a review of Einstein\textquoteright{}s richest laboratory}",
    eprint = "1607.03540",
    archivePrefix = "arXiv",
    primaryClass = "gr-qc",
    doi = "10.1088/1361-6633/aa67bb",
    journal = "Rept. Prog. Phys.",
    volume = "80",
    number = "9",
    pages = "096901",
    year = "2017"
}

@article{Paschalidis:2016vmz,
    author = "Paschalidis, Vasileios and Stergioulas, Nikolaos",
    title = "{Rotating Stars in Relativity}",
    eprint = "1612.03050",
    archivePrefix = "arXiv",
    primaryClass = "astro-ph.HE",
    doi = "10.1007/s41114-017-0008-x",
    journal = "Living Rev. Rel.",
    volume = "20",
    number = "1",
    pages = "7",
    year = "2017"
}

@article{Berti:2002ry,
    author = "Berti, E. and Pons, J. A. and Miniutti, G. and Gualtieri, L. and Ferrari, V.",
    title = "{Are PostNewtonian templates faithful and effectual in detecting gravitational signals from neutron star binaries?}",
    eprint = "gr-qc/0208011",
    archivePrefix = "arXiv",
    doi = "10.1103/PhysRevD.66.064013",
    journal = "Phys. Rev. D",
    volume = "66",
    pages = "064013",
    year = "2002"
}

@article{Kokkotas:1999bd,
    author = "Kokkotas, Kostas D. and Schmidt, Bernd G.",
    title = "{Quasinormal modes of stars and black holes}",
    eprint = "gr-qc/9909058",
    archivePrefix = "arXiv",
    doi = "10.12942/lrr-1999-2",
    journal = "Living Rev. Rel.",
    volume = "2",
    pages = "2",
    year = "1999"
}

@article{Moncrief:1974am,
    author = "Moncrief, V.",
    title = "{Gravitational perturbations of spherically symmetric systems. I. The exterior problem.}",
    doi = "10.1016/0003-4916(74)90173-0",
    journal = "Annals Phys.",
    volume = "88",
    pages = "323--342",
    year = "1974"
}

@ARTICLE{Moncrief:1974an,
       author = {{Moncrief}, Vincent},
        title = "{Gravitational perturbations of spherically symmetric systems. II. Perfect fluid interiors}",
        doi = "10.1016/0003-4916(74)90174-2",
       journal = "Annals Phys.",
        volume = "88",
        pages = "343-370",
        year = "1974"
}

@article{Regge:1957td,
    author = "Regge, Tullio and Wheeler, John A.",
    title = "{Stability of a Schwarzschild singularity}",
    doi = "10.1103/PhysRev.108.1063",
    journal = "Phys. Rev.",
    volume = "108",
    pages = "1063--1069",
    year = "1957"
}

@article{Zerilli:1970se,
    author = "Zerilli, Frank J.",
    title = "{Effective potential for even parity Regge-Wheeler gravitational perturbation equations}",
    doi = "10.1103/PhysRevLett.24.737",
    journal = "Phys. Rev. Lett.",
    volume = "24",
    pages = "737--738",
    year = "1970"
}

@article{Zerilli:1970wzz,
    author = "Zerilli, F. J.",
    title = "{Gravitational field of a particle falling in a schwarzschild geometry analyzed in tensor harmonics}",
    doi = "10.1103/PhysRevD.2.2141",
    journal = "Phys. Rev. D",
    volume = "2",
    pages = "2141--2160",
    year = "1970"
}

@article{Thorne:1980ru,
    author = "Thorne, K. S.",
    title = "{Multipole Expansions of Gravitational Radiation}",
    doi = "10.1103/RevModPhys.52.299",
    journal = "Rev. Mod. Phys.",
    volume = "52",
    pages = "299--339",
    year = "1980"
}

@article{Ivanov:2022qqt,
    author = "Ivanov, Mikhail M. and Zhou, Zihan",
    title = "{Vanishing of Black Hole Tidal Love Numbers from Scattering Amplitudes}",
    eprint = "2209.14324",
    archivePrefix = "arXiv",
    primaryClass = "hep-th",
    doi = "10.1103/PhysRevLett.130.091403",
    journal = "Phys. Rev. Lett.",
    volume = "130",
    number = "9",
    pages = "091403",
    year = "2023"
}

@article{LeTiec:2011ab,
    author = "Le Tiec, Alexandre and Blanchet, Luc and Whiting, Bernard F.",
    title = "{The First Law of Binary Black Hole Mechanics in General Relativity and Post-Newtonian Theory}",
    eprint = "1111.5378",
    archivePrefix = "arXiv",
    primaryClass = "gr-qc",
    doi = "10.1103/PhysRevD.85.064039",
    journal = "Phys. Rev. D",
    volume = "85",
    pages = "064039",
    year = "2012"
}

@article{Detweiler:2008ft,
    author = "Detweiler, Steven L.",
    title = "{A Consequence of the gravitational self-force for circular orbits of the Schwarzschild geometry}",
    eprint = "0804.3529",
    archivePrefix = "arXiv",
    primaryClass = "gr-qc",
    doi = "10.1103/PhysRevD.77.124026",
    journal = "Phys. Rev. D",
    volume = "77",
    pages = "124026",
    year = "2008"
}

@article{LeTiec:2011dp,
    author = "Le Tiec, Alexandre and Barausse, Enrico and Buonanno, Alessandra",
    title = "{Gravitational Self-Force Correction to the Binding Energy of Compact Binary Systems}",
    eprint = "1111.5609",
    archivePrefix = "arXiv",
    primaryClass = "gr-qc",
    doi = "10.1103/PhysRevLett.108.131103",
    journal = "Phys. Rev. Lett.",
    volume = "108",
    pages = "131103",
    year = "2012"
}

@article{Barausse:2011dq,
    author = "Barausse, Enrico and Buonanno, Alessandra and Le Tiec, Alexandre",
    title = "{The complete non-spinning effective-one-body metric at linear order in the mass ratio}",
    eprint = "1111.5610",
    archivePrefix = "arXiv",
    primaryClass = "gr-qc",
    doi = "10.1103/PhysRevD.85.064010",
    journal = "Phys. Rev. D",
    volume = "85",
    pages = "064010",
    year = "2012"
}

@article{Buonanno:1998gg,
    author = "Buonanno, A. and Damour, T.",
    title = "{Effective one-body approach to general relativistic two-body dynamics}",
    eprint = "gr-qc/9811091",
    archivePrefix = "arXiv",
    reportNumber = "IHES-P-98-74",
    doi = "10.1103/PhysRevD.59.084006",
    journal = "Phys. Rev. D",
    volume = "59",
    pages = "084006",
    year = "1999"
}

@article{Damour:2000we,
    author = "Damour, Thibault and Jaranowski, Piotr and Schaefer, Gerhard",
    title = "{On the determination of the last stable orbit for circular general relativistic binaries at the third postNewtonian approximation}",
    eprint = "gr-qc/0005034",
    archivePrefix = "arXiv",
    doi = "10.1103/PhysRevD.62.084011",
    journal = "Phys. Rev. D",
    volume = "62",
    pages = "084011",
    year = "2000"
}

@article{Bini:2012gu,
    author = "Bini, Donato and Damour, Thibault and Faye, Guillaume",
    title = "{Effective action approach to higher-order relativistic tidal interactions in binary systems and their effective one body description}",
    eprint = "1202.3565",
    archivePrefix = "arXiv",
    primaryClass = "gr-qc",
    doi = "10.1103/PhysRevD.85.124034",
    journal = "Phys. Rev. D",
    volume = "85",
    pages = "124034",
    year = "2012"
}

@article{Feng:2021sax,
    author = "Feng, Xuefeng and Lyu, Zhenwei and Yang, Huan",
    title = "{Black-hole perturbation theory with post-Newtonian theory: Towards hybrid waveforms for neutron-star binaries}",
    eprint = "2104.11848",
    archivePrefix = "arXiv",
    primaryClass = "gr-qc",
    doi = "10.1103/PhysRevD.105.104043",
    journal = "Phys. Rev. D",
    volume = "105",
    number = "10",
    pages = "104043",
    year = "2022"
}

@article{Pound:2019lzj,
    author = "Pound, Adam and Wardell, Barry and Warburton, Niels and Miller, Jeremy",
    title = "{Second-Order Self-Force Calculation of Gravitational Binding Energy in Compact Binaries}",
    eprint = "1908.07419",
    archivePrefix = "arXiv",
    primaryClass = "gr-qc",
    doi = "10.1103/PhysRevLett.124.021101",
    journal = "Phys. Rev. Lett.",
    volume = "124",
    number = "2",
    pages = "021101",
    year = "2020"
}

@article{Lindblom:1983ps,
    author = "Lindblom, L and Detweiler, Steven L.",
    title = "{The quadrupole oscillations of neutron stars}",
    doi = "10.1086/190884",
    journal = "Astrophys. J. Suppl.",
    volume = "53",
    pages = "73--92",
    year = "1983"
}

@article{Miao:2023jqe,
    author = "Miao, Zhiqiang and Zhou, Enping and Li, Ang",
    title = "{Resolving Phase Transition Properties of Dense Matter through Tidal-excited g-mode from Inspiralling Neutron Stars}",
    eprint = "2305.08401",
    archivePrefix = "arXiv",
    primaryClass = "nucl-th",
    doi = "10.3847/1538-4357/ad27cd",
    journal = "Astrophys. J.",
    volume = "964",
    number = "1",
    pages = "31",
    year = "2024"
}

@article{Ho:1998hq,
    author = "Ho, Wynn C. G. and Lai, Dong",
    title = "{Resonant tidal excitations of rotating neutron stars in coalescing binaries}",
    eprint = "astro-ph/9812116",
    archivePrefix = "arXiv",
    doi = "10.1046/j.1365-8711.1999.02703.x",
    journal = "Mon. Not. Roy. Astron. Soc.",
    volume = "308",
    pages = "153",
    year = "1999"
}

@article{Xu:2017hqo,
    author = "Xu, Wenrui and Lai, Dong",
    title = "{Resonant Tidal Excitation of Oscillation Modes in Merging Binary Neutron Stars: Inertial-Gravity Modes}",
    eprint = "1708.01839",
    archivePrefix = "arXiv",
    primaryClass = "astro-ph.HE",
    doi = "10.1103/PhysRevD.96.083005",
    journal = "Phys. Rev. D",
    volume = "96",
    number = "8",
    pages = "083005",
    year = "2017"
}

@article{Gualtieri:2001cm,
    author = "Gualtieri, L. and Berti, E. and Pons, J. A. and Miniutti, G. and Ferrari, V.",
    title = "{Gravitational signals emitted by a point mass orbiting a neutron star: A Perturbative approach}",
    eprint = "gr-qc/0107046",
    archivePrefix = "arXiv",
    doi = "10.1103/PhysRevD.64.104007",
    journal = "Phys. Rev. D",
    volume = "64",
    pages = "104007",
    year = "2001"
}

@article{Pons:2001xs,
    author = "Pons, J. A. and Berti, E. and Gualtieri, L. and Miniutti, G. and Ferrari, V.",
    title = "{Gravitational signals emitted by a point mass orbiting a neutron star: Effects of stellar structure}",
    eprint = "gr-qc/0111104",
    archivePrefix = "arXiv",
    doi = "10.1103/PhysRevD.65.104021",
    journal = "Phys. Rev. D",
    volume = "65",
    pages = "104021",
    year = "2002"
}

@article{Lai:1993pa,
    author = "Lai, Dong and Rasio, Frederic A. and Shapiro, Stuart L.",
    title = "{Hydrodynamic instability and coalescence of binary neutron stars}",
    eprint = "astro-ph/9304027",
    archivePrefix = "arXiv",
    reportNumber = "IASSNS-AST-93-22",
    doi = "10.1086/173606",
    journal = "Astrophys. J.",
    volume = "420",
    pages = "811--829",
    year = "1994"
}

@article{Ripley:2023qxo,
    author = "Ripley, Justin L. and Hegade K. R., Abhishek and Yunes, Nicolas",
    title = "{Probing internal dissipative processes of neutron stars with gravitational waves during the inspiral of neutron star binaries}",
    eprint = "2306.15633",
    archivePrefix = "arXiv",
    primaryClass = "gr-qc",
    doi = "10.1103/PhysRevD.108.103037",
    journal = "Phys. Rev. D",
    volume = "108",
    number = "10",
    pages = "103037",
    year = "2023"
}

@article{Kuan:2021jmk,
    author = "Kuan, Hao-Jui and Suvorov, Arthur G. and Kokkotas, Kostas D.",
    title = "{General-relativistic treatment of tidal g-mode resonances in coalescing binaries of neutron stars {\textendash} I. Theoretical framework and crust breaking}",
    eprint = "2106.16123",
    archivePrefix = "arXiv",
    primaryClass = "gr-qc",
    doi = "10.1093/mnras/stab1898",
    journal = "Mon. Not. Roy. Astron. Soc.",
    volume = "506",
    number = "2",
    pages = "2985--2998",
    year = "2021"
}

@article{Saketh:2024juq,
    author = "Saketh, M. V. S. and Zhou, Zihan and Ghosh, Suprovo and Steinhoff, Jan and Chatterjee, Debarati",
    title = "{Investigating tidal heating in neutron stars via gravitational Raman scattering}",
    eprint = "2407.08327",
    archivePrefix = "arXiv",
    primaryClass = "gr-qc",
    doi = "10.1103/PhysRevD.110.103001",
    journal = "Phys. Rev. D",
    volume = "110",
    pages = "103001",
    year = "2024"
}

@article{Creci:2021rkz,
    author = "Creci, Gast{\'o}n and Hinderer, Tanja and Steinhoff, Jan",
    title = "{Tidal response from scattering and the role of analytic continuation}",
    eprint = "2108.03385",
    archivePrefix = "arXiv",
    primaryClass = "gr-qc",
    doi = "10.1103/PhysRevD.104.124061",
    journal = "Phys. Rev. D",
    volume = "104",
    number = "12",
    pages = "124061",
    year = "2021",
    note = "[Erratum: Phys.Rev.D 105, 109902 (2022)]"
}

\end{document}